\documentclass[sigconf,table]{acmart}

\AtBeginDocument{%
  \providecommand\BibTeX{{%
    \normalfont B\kern-0.5em{\scshape i\kern-0.25em b}\kern-0.8em\TeX}}}

\setcopyright{acmcopyright}
\copyrightyear{2020}
\acmYear{2020}
\acmDOI{10.1145/1122445.1122456}

\copyrightyear{2020}
\acmYear{2020}
\setcopyright{acmcopyright}\acmConference[ASE '20]{35th IEEE/ACM International Conference on Automated Software Engineering}{September 21--25, 2020}{Virtual Event, Australia}
\acmBooktitle{35th IEEE/ACM International Conference on Automated Software Engineering (ASE '20), September 21--25, 2020, Virtual Event, Australia}
\acmPrice{15.00}
\acmDOI{10.1145/3324884.3416587}
\acmISBN{978-1-4503-6768-4/20/09}

\usepackage{xspace}
\usepackage{multirow}
\usepackage{cleveref}
\usepackage{xcolor}
\usepackage{booktabs, multirow} % for borders and merged ranges
\usepackage{soul}% for underlines
\usepackage{changepage,threeparttable} % for wide tables
\usepackage[framemethod=tikz]{mdframed}
\usetikzlibrary{shadows}
\usepackage{graphics}
\usepackage{subcaption}
\usepackage{listings}
\usepackage{balance}
\usepackage{graphicx}
\newmdenv[
    tikzsetting= {fill=gray!8},
    skipabove=0.4em,
    skipbelow=0.4em,
    linewidth=1pt,
    innerleftmargin=3pt,
    innerrightmargin=3pt,
    innertopmargin=2pt,
    innerbottommargin=2pt,
    linecolor=gray,
    roundcorner=3pt, 
    shadow=true,
    shadowsize=5pt,
    shadowcolor=gray
]{myshadowbox}

\usepackage{tikz}

\usepackage{listings}
  \lstdefinelanguage{diff}{
    basicstyle=\ttfamily\small,
    morecomment=[f][\color[green]{}],
    morecomment=[f][\color{diffincl}]{+\ },
    morecomment=[f][\color{diffrem}]{-\ },
  }

\newcommand{\rom}[1]{\uppercase\expandafter{\romannumeral #1\relax}}

\newcommand{\etal}{\hbox{\emph{et al.}}\xspace}
\newcommand{\eg}{\hbox{\emph{e.g.,}}\xspace}
\newcommand{\ie}{\hbox{\emph{i.e.,}}\xspace}
\newcommand{\wrt}{\hbox{\emph{w.r.t.}}\xspace}

\newcommand{\train}{{\sc training}\xspace}
\newcommand{\heldout}{{\sc held-out}\xspace}
\newcommand{\bo}{\textit{b}\xspace}
\newcommand{\po}{\textit{p}\xspace}
\newcommand{\bs}{$\tilde{b}$\xspace}
\newcommand{\ps}{$\tilde{p}$\xspace}

\newenvironment{result}
{\begin{myshadowbox}}
{\end{myshadowbox}}
\begin{document}

\title{Patching as Translation: 
the Data and the Metaphor}
%The Name of the Title is Hope
%}

\author{Yangruibo Ding}
\affiliation{%
  \institution{Columbia University}
  %\city{New York City}
  %\state{New York}
  %\country{USA}
}
\email{yangruibo.ding@columbia.edu}

\author{Baishakhi Ray}
\affiliation{%
  \institution{Columbia University}
  %\city{New York City}
  %\state{New York}
  %\country{USA}
}
\email{rayb@cs.columbia.edu}

\author{Premkumar Devanbu}
\affiliation{%
  \institution{University of California, Davis}
  %\city{Davis}
  %\state{California}
  %\country{USA}
}
\email{ptdevanbu@ucdavis.edu}

\author{Vincent J. Hellendoorn}
\affiliation{%
  \institution{University of California, Davis}
  %\city{Davis}
  %\state{California}
  %\country{USA}
}
\email{vhellendoorn@ucdavis.edu}

\newcommand{\todo}[1]{{\color{red}{\small\bf #1}}}

\newcommand{\pd}[1]{{\footnotesize \todo{PD:  {\color{blue} #1}}}}
\newcommand{\brc}[1]{{\footnotesize \todo{BR:  {\color{blue} #1}}}}
\newcommand{\rdc}[1]{{\footnotesize \todo{RD:  {\color{blue} #1}}}}
\newcommand{\vhc}[1]{{\footnotesize \todo{VH:  {\color{blue} #1}}}}

\begin{abstract}
Machine Learning models from other fields, like Computational Linguistics, have been transplanted to Software Engineering tasks, often quite successfully. Yet a transplanted model's initial 
success at a given task does not necessarily mean it is well-suited for the task. In this work, we examine a common example of this phenomenon: the conceit that ``software patching is like language translation''.
We demonstrate empirically that there are subtle, but critical distinctions between sequence-to-sequence models and translation model: while program repair benefits greatly from the former, general modeling architecture, it actually suffers from design decisions built into the latter, both in terms of translation accuracy and diversity. 
Given these findings, we demonstrate how a more principled approach to model design, based on our empirical findings and general knowledge of software development, can lead to better solutions.
Our findings also lend strong support to the recent trend towards synthesizing {\em edits} of code conditional on the buggy context, to repair bugs. 
We implement such models ourselves as "proof-of-concept" tools and empirically confirm that they behave in a fundamentally different, more effective way than the studied {\em translation-based} architectures.
%We propose several models that leverage the same machine learning tools, but whose architecture, data presentation, and metrics are specialized for the software engineering task. The resulting models perform significantly better than the studied baseline, especially in more program repair appropriate metrics. 
Overall, our results demonstrate the merit of studying the intricacies of machine learned models in software engineering: not only can this help elucidate potential issues that may be overshadowed by increases in accuracy; it can also help innovate on these models to raise the state-of-the-art further. We will publicly release our replication data and materials at \url{https://github.com/ARiSE-Lab/Patch-as-translation}.
\end{abstract}





\maketitle
\section{Introduction}
\label{sec:intro}
Recent work has applied a wide variety of machine learning models to practical software engineering tasks, including code completion, automated program repair, and code comment generation. These models excel at learning general patterns from large amounts of diverse data, even when training data is relatively unstructured. This combination enables one to 
%makes it convenient to 
simply \emph{transplant} successful models from related fields, \eg from computational linguistics, to software engineering. Yet, even if these models provide reasonable performance, the transplanted model may still not be appropriate for the task; many of these models were designed for paradigms that differ subtly, yet significantly.

In this work, we conduct a systematic empirical case-study to illustrate how transplanted models can fail in the targeted task domain, focusing specifically on the concept of ``patching as translation'' as a typical example of this phenomenon. A range of recent work has adopted neural machine translation (NMT) models to learn to repair programs by ``translating" the buggy code to the repaired code~\cite{tufano2018nmt_bug_fix, chen2018sequencer, chakraborty2018codit, lutellier2020coconut}. We argue that there are three general concerns with this type of approach, and show concretely how these manifest in ``patching as translation'' through empirical analysis:

\vspace{-1.5mm}
\paragraph{Task design:} Deep Learning (DL) models transform their inputs into a compact set of features that stores the important information, which it then uses to produce the required target. A wide range of DL architectures have been proposed that do so, but regardless of the specific architecture or task, it is self-evident that all the relevant information needed to generate the target must already exist in the input. While that is (largely) a fair assumption for natural language translation, where we can assume that the input \& output sentences express the same idea, it is questionable for source code repair: we show evidence that buggy fragments often lack the information required to repair them. Reliably choosing the correct repair may even be impossible without access to a very broad context (including surrounding files), in the absence of which this task is inevitably ambiguous for many real-world bugs.

\vspace{-1.5mm}
\paragraph{Architectural design:} Given a task where deep learning is feasible, one must choose a model architecture that supports the transformation from input to output, in as realistic and simple a manner as possible. This is done by ensuring that prior knowledge of the task (including dependencies and structural properties) are built into the model design.
Architectures for machine translation rely heavily on the auto-regressive nature of text: language is generally produced one word (or token) at the time in left-to-right manner, \eg in speech or writing; the standard NMT encoder-decoder architecture generates translations correspondingly. While this works very well for NMT, its relevance to practical program repair is tenuous at best: empirically, many repairs just copy (nearly all) tokens from the buggy line, with very few changed tokens (often just one). As such, both bug and patch share a large identical prefix, but the difference in the subsequent tokens is crucial. We demonstrate that models struggle to predict this transition, as the large amount of copying distracts (and inflates) the training quality signal.

\vspace{-1.5mm}
\paragraph{Objective design:} finally, models are trained by computing a loss for their predictions relative to
a ``gold" output, using a \emph{loss function}. This function is usually a \emph{differentiable} proxy for the actual quality of the model, because such qualitative assessments tend not to be differentiable. In machine translation, the training loss is usually based on the probabilities of the correct token;
the actual quality of the trained model is measured with BLEU scores (or the like) that measure overlap between the generated and ground-truth translation. However, such overlap measures are inappropriate for program repair. For reasons stated earlier (few token changes, lack of contextual information), the quality of a produced repair often correlates very poorly with the number (and placement) of tokens it shares with the desired output. For instance, the trained model emits many syntactically incorrect repairs, as well as many very similar patches for a given bug, rather than exploring a range of alternatives. This yields poor performance in a search-based setting in which they are popularly used (given failing test cases).

Having studied these concerns empirically, we provide strong evidence that the metaphor of ``patching as translation" is inappropriate for the task. On the other hand, our results shed light on the effectiveness of a recent competing approach: repairing bugs by synthesizing \emph{edits} of code conditional on the buggy context~\cite{Dinella2020HoppityLG, vasic2019pointernet, tarlow2019graph2diff}. Not only does this approach agree better with our empirical findings, we implemented a general form of this architecture, which predicts insertions and deletions relative to the bug rather than the entire patch, and show that this change in particular produces a model that both generates the correct patch more often, and provides better sampling behavior (\ie higher top-K prediction accuracy using beam search). We also implemented a generic contextual extension (compatible with both models) to assess whether our empirical findings of the importance of context could be integrated easily into the models; this enhancement proved less effective, likely because existing models struggle to model large windows of context. This finding highlights that empirically observing an issue and effectively addressing it are not always the same; we leave this challenge for future work.

\section{Background}
\label{sec:background}
To study the transplanting of architectures from neural machine translation (NMT) to automated program repair, we need to understand both domains. In this section, we first discuss NMT -- its conceptual needs and corresponding architectural designs, and then Automated Program Repair -- its empirical characteristics and practical use.

\subsection{Neural Machine Translation}
NMT aims to convert an expression in a source language (\eg English) into a semantically equivalent expression in a target language (\eg French). This is generally both quite feasible and fairly deterministic: a given English sentence almost certainly has a French translation that is both a good French expression on its own and preserves all the information in the original English expression (\emph{i.e.,} it could be translated back to a comparable English phrase).\footnote{In practice, context is sometimes required, \eg to determine if an expression is meant sarcastically, which may alter its translation. There can also be multiple valid translations for one expression (\eg literal vs. idiomatic). Even so, generated translations that overlap strongly with the ground-truth are rated highly by human translators \cite{papineni_02_bleu}.}

A natural fit for this task is the {\em encoder-decoder} architecture, which consists of two components: 1) an encoder that learns to compactly encode the important information from the source language expression, and 2) a decoder, which transforms that information into an equivalent expression in the target language. Encoder-decoder style models can address many types of transformations between two domains (\eg from image to textual description) and are typically instantiated with specific encoder-decoder architectures for a given problem that reflect some knowledge about that problem's domain.
This simplifies the otherwise complex task of representing and producing a very wide range of inputs. For example, in computational linguistics, sequence-to-sequence {\em (seq2seq)}~\cite{sutskever2014seq2seq} models are a well-established way to generate text one token at a time, in a left-to-right manner -- this linear order reflects how language is often generated in speech and writing. Seq2seq models exploit this structure by both representing and generating expressions with a strong emphasis on the left-to-right relations between tokens; especially in the decoder component, which (in nearly all popular models) produces tokens ``auto-regressively", meaning that tokens are produced one by one, and every previously generated token is fed back to the model to produce the next token. 
% This greatly simplifies the learning task, as at training time we can simply provide each correct token regardless of the model's output, which enables more smooth, stable training.

\vspace{-2mm}
\paragraph{Practical seq2seq models}
{\em Seq2seq} models have achieved great success in the NMT field. Recurrent Neural Networks (RNNs) were popular for many years, but had difficulties in remembering long-term dependencies. In an RNN, all the information of a source sentence is encoded into a hidden state from left to right; the final hidden state is then passed to the decoder, which attempts to reconstruct the target expression from this information. This puts inordinate strain on that single hidden state, which tends to cause the model to forget tokens seen long ago. Long short-term memory (LSTM)~\cite{hochreiter_1997_lstm} and gated recurrent unit (GRU)~\cite{Cho2014OnTP} were introduced to mitigate this bottle-neck by better separating long-term and token-specific information, and did significantly improve the performance of RNN-based NMT models, but ultimately suffered from the same concerns. Attention-based mechanisms \cite{Bahdanau2015NeuralMT} were introduced to allow the decoder to ``attend" to any given intermediate state from the encoder (rather than only the final one), which greatly improved performance.
%\vhc{citation for attention; probably just Bahdanau 2014}
Most recently, the Transformer~\cite{vaswani2017transformer} model generalized this idea to relying entirely on attention mechanisms to both encode inputs and generate outputs. The Transformer model proposes multi-headed (self-)attention interspersed with feed-forward networks that enables both encoder and decoder to attend to any set of tokens across arbitrarily long distances. These models are also highly parallelizable. We adopt this model in our work.

\vspace{-2mm}
\paragraph{Seq2seq models in SE}
Hindle \etal observed that source code is ``natural" ~\cite{Hindle2012naturalness}, \emph{viz.,} with strong local dependencies similar to natural languages like English.  Many language models have been applied to software engineering tasks. More recently, this includes a range of applications of the {\em seq2seq} architecture in modeling source code. Existing work has exploited their potential in several SE tasks, such as code summarization~\cite{iyer2016summarizing}, code migration~\cite{gu2017deepam} and program repair~\cite{tufano2018nmt_bug_fix, chen2018sequencer, chakraborty2018codit, lutellier2020coconut}.
The prevalent approach is to treat source code as a sequence of tokens with implicit or explicit structures (\eg abstract syntax trees) \cite{chakraborty2018codit}. The encoder learns the distribution of such structured language, which is then translated into the target domain, either program languages (PL) or natural languages (NL).
Existing work has especially spent effort on learning meaningful code representations to adapt {\em seq2seq} models for source code modeling~\cite{allamanis18learning, Alon2019code2vec, zhang2019astnn}; these approaches focus on encoding rich characteristics of programming languages, besides just the tokens. Our work takes a more general view, and aims to study the feasibility of constructing good tools based on both the problem design and general model architecture, which is largely orthogonal to learning better representations of code. We recognize that researches in both directions are necessary and significant to improve the efficacy of deep learning approaches in SE.

\vspace{-2mm}
\paragraph{Relevance of Models to Tasks.}
All these models excel at learning generalizable patterns from large amounts of diverse data and are \emph{prima facie} at least somewhat applicable to source code, to the extent that it reflects natural language characteristics. However, different tasks come with their own concepts and peculiarities, and the models should reflect the phenomena specific to the task. For example, code summarization and code migration are more like NL translation tasks, since both their goals are to encode and preserve the semantics of their inputs (code fragments), just in different vocabularies (concise natural language, and another code context respectively).
On the other hand, software engineers behave differently when repairing a program. Developers tend to fix a buggy fragment by making minor changes rather than entirely rewriting it. Furthermore, the semantics of the buggy fragment are by definition not preserved; the express goal is to introduce semantically new content (and possibly remove some) so as to change the meaning of a fragment. None of this disqualifies the use of seq2seq models \emph{per se}, but its built-in assumptions should at least be carefully evaluated empirically, and, if necessary, its application should be changed to better reflect the domain.

\vspace{-2mm}
\subsection{Automated Program Repair}
Automated program repair (APR) is a task of keen interest in SE. The aim is to fix software bugs with minimal human intervention. Classic APR techniques can be categorized into 1) generate-and-validate (G\&V) \cite{Goues2012GenProgAG, qi_2015_kali, qi_2014_rsrepair, Simfix:2018} or 2) synthesis-based approaches~\cite{le_2017_s3}. G\&V approaches automatically generate patches and validate the candidates using a set of test cases that reveals the bug. To generate fixes, one effective approach is to mutate (\eg insert, replace) the buggy code according to code snippets in the current project that occur in similar contexts \cite{Simfix:2018}. Synthesis-based approaches create constraints that satisfy all test cases, and then solve them and produce patches from the solutions.

\noindent
\paragraph{NMT for APR.}
Tufano ~\etal~\cite{tufano2018nmt_bug_fix} proposed to use machine translation to repair programs and empirically studied the feasibility of translating buggy programs into fixed ones. They applied multi-layer RNNs with either LSTM or GRU nodes to predict patches of abstracted, real bugs, and report promising performance. Chen ~\etal~\cite{chen2018sequencer} subsequently introduced {\em SequenceR}, an end-to-end framework to repair one-line Java bugs. They used NMT models to learn the implicit bug-repair patterns by training the model with ~35k bug-fix pairs. Besides the buggy line itself, they also considered code context to allow long-range dependencies in fixes; they include the entire class in which the bug is located, which they {\em abstract} to reduce the input size. CODIT~\cite{chakraborty2018codit} developed a tree-based NMT model to produce code edits and bug fixes. It first translates the tree structure of code and utilizes the structural information to assist the generation of code tokens. CODIT also includes the tree nodes around the bug as context to predict meaningful patches. CoCoNuT~\cite{lutellier2020coconut} ensembles multiple NMT models to capture diverse fix patterns. The authors argue that incorporating context is essential for fixing bugs, so they apply a separate encoder specifically for learning the buggy context. 
%yet ineffective for deep learning models, so they ignore the buggy context.

Although these existing works apply a wide range of models, they all treat program repair as a translation task; these tools encode a limited program window around a bug and learn to transform it to repaired code based on historical repairs. The premise is that translation is both a suitable model and that the buggy code (with its context) provides sufficient information to succeed. It is thus past time to ask the following, high-level question, which has yet to be addressed from an empirical perspective:

\vspace{1.5mm} %5mm vertical space
\noindent \textbf{RQ1.} \textit{Is it generally feasible to translate buggy programs to repairs?}
\vspace{1.5mm} %5mm vertical space

While these approaches all indicate that Seq2Seq learning holds promise
for learning patterns of transformation between bugs and patches, they struggle to outperform many G\&V tools that applied human-designed rules to fix defects. One explanation is that the search space of repairs is prohibitively large~\cite{fan2016analysis}, among others due to the large and highly local vocabulary and patterns endemic to software \cite{hellendoorn2017deep}, as well as the length of buggy fragments. Intuitively, however, the search space need not be so large at all: in real-world development, modifications made to code during repair are mostly small, limited to a few tokens rather than completely reconstructing a statement. Is the reliance on translation putting models at a disadvantage by artificially expanding the search space?
%To successfully empower the machine to generate meaningful patches, model designers need to better understand the fixing behaviors performed by human developers and create architectures properly.
It is again worth determining this empirically, by asking:

\vspace{1.5mm} %5mm vertical space
\noindent \textbf{RQ2.} \textit{Do machine translation architectures mischaracterize real-world fixing behavior, and does this disadvantage their performance?}
%\vspace{1.5mm} %no vertical space since a paragraph starts next

\vspace{-2mm}
\paragraph{Deep Learning for APR.}
Besides translating buggy code to fix it, recent work has proposed deep learning models that learn to specify the buggy locations that need to be modified together with the edits to be made. DeepFix \cite{gupta2017deepfix} implemented a {\em seq2seq} attention network to fix compiler errors. As input, the program is represented as a sequence of (line number, tokens) pairs, and the model predicts a single (buggy line number, patch) pair as a repair. Vasic~\etal~\cite{vasic2019pointernet} proposed using pointer networks \cite{vinyals2015pointernet} to jointly learn to localize and repair a specific class of bugs known as \textsc{VarMisuses}. Their network jointly predicts two output ``heads", one to locate the buggy token and one for its replacement. Tarlow~\etal~\cite{tarlow2019graph2diff} introduced an edit-based model called {\em Graph2Diff} that uses a graph neural network as an encoder and a Transformer as decoder. This model transforms a program graph into a {\em ToCoPo} sequence of AST edits that transform the buggy program into a repaired version.

By directly learning the locations of incorrect tokens and the edits to be made, these edit-based methods provide an approach to learning bug fixing with a very different \emph{loss}, which is not trivially reduced by maximizing the token overlap between the bug and repair. The impact of this loss can be substantial in determining the kind, and robustness, of local minima that the neural network finds during training. We thus implement a simple version of this model ourselves to empirically study the impact of the objective function on both our baseline model and this edit-based model. This way, we study the impact of the objective function on the models by studying \emph{the models' results} themselves, asking the following:

\vspace{1.5mm} %5mm vertical space
\noindent\textbf{RQ3.} \textit{How well does the NMT objective function apply to Automated Program Repair?}

\vspace{-2mm}
\section{Methodology}
\label{sec:methodology}
The goal of this work is to provide an empirical and conceptual analysis of 
the relevance of deep learning models (originally developed for NMT)
in SE contexts. 
As such, we emphasize that it is \emph{not our goal} to produce a state-of-the-art bug detector, or replicate prior work. Rather, we identify a general, representative approach (seq2seq for program repair), that reflects a direct adoption of models from a related field to SE tasks, and study its limitations. Naturally, prior work has covered a wide range of applications and modifications of this method, and may be immune to some of our findings, but this does not discount the general result of our analysis: that adapting
deep-learning models designed for other fields to SE 
requires a principled, empirically and conceptually grounded approach. 

\vspace{-2mm}
\subsection{Scope}
Concretely, we focus on a relatively simple form of automated program repair in which we translate a given buggy line to its repaired counter-part. We thus assume that we have the bug already localized and that it is confined to exactly a single line. This is the most direct form of ``repair as translation'', in which an off-the-shelf translation model is used on two software ``sentences": the buggy version and the repair.

\vspace{-2mm}
\subsection{Data}
We collect our bugs from the history of the 10,235 most-starred Java repositories on Github on March 30th, 2020. We analyzed each project's entire commit history and extracted any commits that altered precisely a single line in a single Java file, disregarding any (spurious) changes to whitespace. We then compared the corresponding commit messages against a relatively simple keyword-based check~\cite{Ray2016buggycode} to heuristically find commits labeled as \eg ``fix" or ``bug". We note that, although this heuristic is not particularly precise, the characteristics we found in our data were very similar between those marked as fixes and other one-line changes, so we expect this to have little impact on our analysis. This process resulted in \emph{ca.} 60,000 bug fixes across 8,644 projects in our dataset. In the course of our analysis of this data, we manually checked a number of the collected samples and confirmed that the vast majority of these were indeed bug fixes.

\vspace{-2mm}
\subsection{Experiment Setup}
Given the collected dataset, we first analyze the characteristics of real fixes and then train NMT models on these samples to predict patches. To answer our research questions, we study both characteristics of the real-world bug-fixing behaviors and of the model-generated patches.

\subsubsection{Bug context.} 
\label{subsubsec:buggy_context}
We design experiments to explore the importance of a bug's lexical context when fixing defects. In natural languages, context (the text surrounding an expression) has a direct effect on the way people understand a specific expression and can help avoid ambiguity in communication. Similarly, the context of a buggy line is the code surrounding, as in, both preceding and succeeding, the bug. This can variously be chosen to include up to $N$ lines of code above and below the bug, the surrounding function, or even the whole file (or project). This context can provide vital information (\eg variable definitions, conditional statements) for understanding the defect and the necessary repair. We study the role of variously sized contexts for both disambiguation and providing necessary information in \cref{subsec:semantics}.

%\vspace{-2mm} Let's not remove too much whitespace between sections; it looks congested.
\subsubsection{Similarity analysis.}
\label{subsubsec: similarity_metrics}
We noted earlier that software engineers tend to make small changes when fixing defects, likely because bugs correspond to only minor flaws in the code, and perhaps also because making few changes reduces the risk of introducing new bugs. Given this, we evaluate the similarity between real bugs and patches empirically across three similarity metrics: 

(i) {\em Edit distance} (precisely, {\em Levenshtein distance}) is a metric that quantifies the difference of two sequences by the minimum number of edits (deletions, insertions, or substitutions) required to transform one into the other. 

(ii) {\em Jaccard similarity} (effectively intersection-over-union) calculates the ratio of overlapping n-grams between two sequences divided by their union. Jaccard similarity is usually just applied to token-level similarity; we extend it to the average of 1 through 4-gram overlap to better capture both token-level similarity and matches in their ordering.

(iii) {\em Bilingual evaluation understudy} (BLEU) is popularly used to evaluate the quality of machine translations, and is considered to have a high correlation with human assessments of similarity~\cite{papineni_02_bleu}. This is an asymmetric measure that captures how similar the model prediction (the ``hypothesis") is to the ground-truth translation (the ``reference"). BLEU also counts n-gram matches between the prediction and the ground-truth, and normalizes these w.r.t. the predicted sequence length, which causes the asymmetry.

These three metrics above will be used frequently across our analysis to measure similarity in different aspects.

\subsubsection{Model training and BPE}
\label{sec:training}
To inspect the performance of NMT models on program repair, we trained and evaluated a vanilla Transformer model~\cite{vaswani2017transformer} on our dataset. We split the whole dataset into three parts across organizations, {\sc train/valid/test}, with a ratio of 90\%:5\%:5\%. We trained the model on the ca. 55K bugs in the {\sc train} set, optimized it for held-out performance relative to the {\sc valid} set and finally evaluated the performance on {\sc test} set.

In the field of natural language processing, {\em Byte-pair-encoding} (BPE)~\cite{sennrich-etal-2016-neural} is a widely used method to encode rare and long words into frequent sub-tokens; this way, tokens that were not seen in full during training can still be predicted accurately at inference time. BPE splits a word (\eg "coding") into a list of more frequent sub-tokens (\eg ['cod', 'ing']). In programming languages, vocabulary innovation is even more rampant, as developers tend to name a variable or method using a combination of words (\eg \texttt{isNullOrEmpty}) \cite{hellendoorn2017deep}. Karampatsis~\etal~\cite{karampatsis2020big} show that BPE can effectively address this issue in big code applications, so we apply this to our model input as and predictions as well.

%\vhc{TODO: explain training, BPE, BLEU, context, etc; shouldn't be explained in section 4}

%\section{Repair As Translation}
\vspace{-2mm}
\section{Analysis}
\label{sec:translationrepair}
In this section, we empirically analyse the characteristics of program repair on real-world bug fixes with a joint focus on the relation to natural language translation and on factors that influence accuracy for program repair. In particular, we study the characteristics of the ground-truth data (\ie the real bugs and patches with their context), and of the patches generated by our NMT model, as trained in \Cref{sec:training}. For the rest of this section, 
%we follow the organization presented in~\Cref{sec:intro}:
we scrutinize the adequacy of translation as a paradigm in \Cref{subsec:semantics},
%\bray{use a different section heading, right now the heading is a bit confusing}
we identify architectural concerns in \Cref{subsec:architecture}, and we quantify their impact on model performance in \Cref{subsec:objective}.

\vspace{-2mm}
\subsection{Task Design}
\label{subsec:semantics}
Translation, as a task, is intended to facilitate communication across language barriers. Hence, by design, it must preserve the semantics between the source language, and the target language\textemdash any translation that changes the semantics is unacceptable.
By contrast, in program repair, the semantics of source and target are meant to differ, as the buggy version contains incorrect program behavior that the fixed version is supposed to correct. To do that, engineers deliberately change (add/delete/replace) incorrect tokens with correct ones. Imitating such changes with a machine learner is non-trivial, especially since the learner usually only has access to the bug and the fix, but not the knowledge latent in the developer's mind to reason about the fix.

For instance, the fix may introduce, \emph{de novo}, tokens that are not in the buggy lines, \eg a new API call. In such cases, a model has to learn to pick those new tokens from across its entire known vocabulary.  If the replacement is a common fix pattern, this might be easy enough to learn; otherwise, this leads to a vast search space of candidate repairs. The latter case is common enough; developers often use methods of their own creation, defined in adjacent files, or string or numerical patterns specific only to that project. It is thus important to \emph{quantify}, even just approximately, how much information the model needs and how much it has access to from its training data, which tends to comprise the buggy lines and an optional window of code context.  
Although it is non-trivial to inspect the learned ``black-box" model and extract what it infers about a given buggy line, we can identify the \textbf{gaps} between "what program repair needs" and "what machine translation can supply".

%\vspace{-1mm}
\subsubsection{Program repair needs (lots of) contextual information}
\label{subsubsec:new_vocab}
Patches that introduce new vocabulary (relative to the buggy line) require the model to conjure up novel tokens, \emph{ex nihilo}. Given that code vocabulary is highly diverse and often strongly specific to a given project, package and file \cite{hellendoorn2017deep}, doing so from the buggy line alone may require an unreasonable level of ingenuity from the model. \Cref{tab:new_vocab} quantifies this: first, nearly 90\% of patches introduce new vocabulary relative to their buggy source, which is true regardless of sub-tokenization (even using the BPE approach). Furthermore, these are not at all just typical program tokens or local variables; we paired the buggy line with increasing windows of context (explained in \Cref{subsubsec:buggy_context}) and find that the unseen tokens introduced by the patch are still rarely borrowed from any immediate buggy context; they are sometimes present in the file as a whole, but in locations far away from the bug. Nearby tokens are a bit more likely to share some sub-token(s) with the patch, but rarely provide the entire missing link. Given that modeling large volumes of code (\ie many hundreds, or thousands of tokens) at once is often prohibitively expensive for current deep learned models, this can seriously affect models that incorporate only modest levels of context, such as the surrounding few lines or function.

\begin{table}[!htb]\centering
\vspace{-2mm}
\caption{\small \textbf{Ratio of patches with new vocabulary relative to the buggy snippet given a context window that ranges from \textit{None} (\ie only buggy lines -- a typical translation setting), to a given number of lines (symmetrically around the bug), and finally to the entire file.}}
\begin{tabular}{llr}\toprule
\small
\multirow{2}{*}{Context included} &\multicolumn{2}{c}{Patches introducing unseen tokens} \\\cmidrule{2-3}
& without BPE & with BPE \\\midrule
None & 89.5\% & 86.2\% \\
10 lines & 73.0\% & 64.2\% \\
20 lines & 68.7\% & 59.9\% \\
Whole file & 49.5\% & 38.6\% \\
\bottomrule
\end{tabular}
\label{tab:new_vocab}
\vspace{-2mm}
\end{table}
This is not merely a matter of richer training data either; a large proportion of project-specific tokens are not found in any other projects \cite{hellendoorn2017deep}, so it is quite unlikely that our model would have seen many of these at training time. 
We note that this is in contrast to other paradigms of program repair, such many G\&V models, which instead \emph{search} for patches from across many surrounding files, rather than aim to encode a context directly into a translation.

\subsubsection{Without context, program repair is inherently ambiguous}
\label{subsubsec: ambiguity_without_context}
As discussed, a learner would certainly struggle to capture enough information from the buggy program alone. Fortunately, these learners are equipped with the capacity to transfer many insights from their training data to new examples. Perhaps they can predict the missing semantic information from those bug-repair pairs?

Although it is again impossible to quantify what the model can do, we also again argue that it is perfectly sound to lower-bound its potential by estimating how much of the requisite information it has access to at training time. Concretely, the training data contains many ``similar" bugs to those seen at test time (which we will quantify in various ways), so the model might learn the transformations that produced patches from those similar bugs and apply the same insight, \eg to predict the missing vocabulary. But, this is contingent on similar bugs indeed producing similar repairs; if repairs for similar bugs routinely diverge, then the model is reasoning about highly ambiguous data and will have to learn a wide range of valid transformations for a single defect in the training data.

To simplify this discussion, let \bo be a bug in the \heldout portion of our dataset and \po be the patch of \bo~. Assuming we had some oracle that can provide ``similar" bugs \bs (with patch \ps) for \bo, specifically from our \train data, we would ideally expect information about the relative change needed to repair \bo to be transplanted from \ps. If that is generally true, then our model can learn similar transformations for similar bugs and thereby generate new vocabulary and patterns that are not present in the buggy context.

\begin{figure}
  \begin{subfigure}{4.1cm}
    \includegraphics[width=4.1cm]{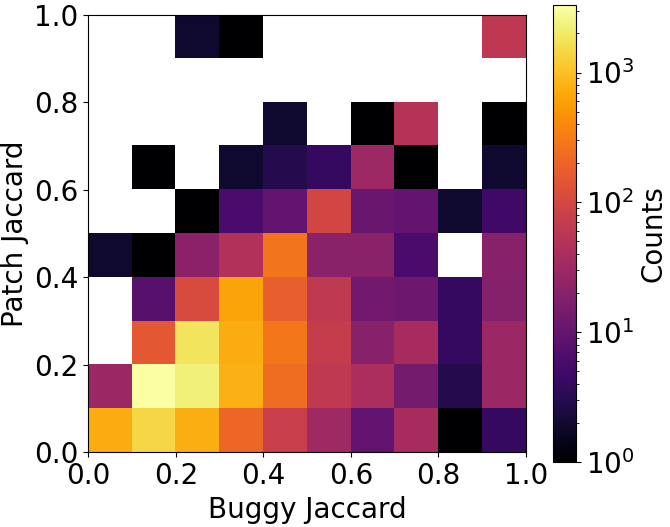}
    \caption{Intersection-over-union}
    \label{subfig:ambiguity_jaccard}
  \end{subfigure}
  \begin{subfigure}{4.1cm}
    \includegraphics[width=4.1cm]{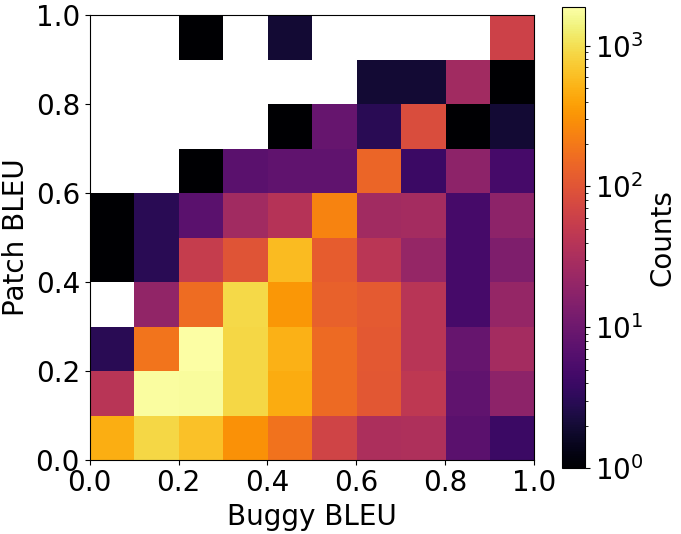}
    \caption{BLEU}
    \label{subfig:ambiguity_bleu}
  \end{subfigure}
  \caption{\small Correlation between bugs similarity and patches similarity. X-axis indicates the bug/similar-bug similarity, and y-axis indicates the patch/similar-patch similarity. The corresponding count of each grid is normalized on a log scale.}
  \label{fig:semantic_space}
  
  %\bray{increase the font of the labels, also mention it is in log scale}
\vspace{-6mm}
\end{figure}

To quantify this, we find the top-3 most similar bugs for each \bo in the \heldout using 4-gram Jaccard index (see \cref{subsubsec: similarity_metrics}), which we label $\tilde{b}_1$, $\tilde{b}_2$, $\tilde{b}_3$, among bugs in the \train data.
We then extract the corresponding \po~ and $\tilde{p}_i$ and evaluate the patch similarity SIM[\po, $\tilde{p}_i$] in relation to the bug similarity SIM[\bo, $\tilde{b}_i$]. To facilitate transferring repair patterns, we should hope that similar bugs produce similar repairs. We visualize this as a heat map (\Cref{fig:semantic_space}) to show the correlation between bugs' similarity and patches' similarity: for each grid in the heat map, we count the number of samples with SIM[\bo, $\tilde{b}_i$] and SIM[\po, $\tilde{p}_i$] in the corresponding range, and then color the grid based on these counts. To make the color contrast more identifiable, we log-normalize the counts.

We calculate the \bo $\to$ \bs and \po $\to$ \ps similarity scores using both the Jaccard index and BLEU scores; the latter is more appropriate for translation because it is asymmetrical, capturing the overlap from the perspective of the translation target. This aligns well with the task's directionality: we want to quantify what information is transferred from \train to \heldout, not vice versa. The result is shown in  \Cref{fig:semantic_space}; both metrics yield a similar pattern: bug-similarity only partially correlates with patch-similarity. Both graphs show a ``smeared out" pattern in which similar bugs tend to produce patches with typically less similarity, rather than a strongly pronounced diagonal, that would indicate that patches relate to one another as their bugs do. Worse, many bugs have only neighbors with low similarity to begin with. These lower scores tend to just reflect spurious overlap due to the large portion of ``closed-vocabulary" tokens (\eg brackets, keywords) in source code, which is also evident from the main hotspot being at (0.25, 0.25).

We are particularly interested in pairs that share a relatively large number of tokens and patterns; \ie those with similarity scores greater than 0.5. For example, the code: ``\texttt{private boolean isName = false;}" and code: ``\texttt{private boolean isName = true;}" yield a BLEU-score of 0.57, and they indeed look alike (only differ in boolean value). 
If similar bugs are (predominantly) fixed in similar ways, then we should expect that to translate into high patch similarity, which would allow the model to copy the appropriate repair patterns. Unfortunately, \Cref{tab:semantic_space}, which breaks down the highly similar bugs specifically, paints a different picture: here too, the similarity between bugs has nearly no discernible relation to that of their patches. Even highly similar bugs' patches do not score above 0.5 half the time, which is actually lower than their respective bugs. For instance, a common bug in our dataset, "\texttt{LOG.error(e);}", presents with many dissimilar patches including "\texttt{LOG.warn(e);}" and "\texttt{LOG.error(``Can't read settings for " + tool, e);}". The BLEU score between these two patches is just 0.12, and we can tell that this bug was fixed with very different intentions. In other words, relying on similar bugs to transplant patch information is almost entirely ineffective.

\begin{table}[t]\centering
%\vhc{TODO: make heatmap}
%\vspace{-3mm}
\caption{\small Similar bugs do not always have similar patches.}\label{tab:semantic_space}
\footnotesize

\begin{tabular}{ccccccc}\toprule
\multirow{2}{*}{bug/similar-bug BLEU} &\multirow{2}{*}{\# Samples} &\multicolumn{4}{c}{patch/similar-patches BLEU} \\\cmidrule{3-6}
& &\multicolumn{2}{c}{$<$ 0.5} &\multicolumn{2}{c}{$\geq$ 0.5} \\\cmidrule{1-6}
$\geq$ 0.5 &2038 &\multicolumn{2}{c}{66.0\%} &\multicolumn{2}{c}{34.0\%} \\\midrule
$\geq$ 0.6 &1143 &\multicolumn{2}{c}{62.3\%} &\multicolumn{2}{c}{37.7\%} \\\midrule
$\geq$ 0.7 &561 &\multicolumn{2}{c}{54.2\%} &\multicolumn{2}{c}{45.8\%} \\\midrule
$\geq$ 0.8 &258 &\multicolumn{2}{c}{46.1\%} &\multicolumn{2}{c}{53.9\%} \\\midrule
1 &173 &\multicolumn{2}{c}{49.1\%} &\multicolumn{2}{c}{50.9\%} \\
\bottomrule
\end{tabular}
\vspace{-5mm}
\end{table}

This demonstrates a substantial \emph{inherent ambiguity} in program repair based on just a buggy line (though not necessarily to program repair in general): for a given bug, the learned program repair history provides a mixed signal of many candidate repairs with distinct semantics.
This matches our intuition as well: just how a given fragment is buggy, and what specific repair among many valid semantic transformations is appropriate depends on a vast array of factors, many of which are not enshrined in the code at all (\eg project requirements, developer preferences), let alone the buggy line (or even function) itself.

%Thus, we propose the second \textbf{gap}:
%\begin{result}
%Program repair has \emph{inherent ambiguity}, so that NMT models will get puzzled when learning fix patterns from bug-fixing history.
%\end{result}

%\subsubsection{Perhaps the models is smarter?}
\subsubsection{The challenges of new vocabulary}
Finally, it might still be feasible for the model to ``guess" at novel tokens and break the ambiguity if they can be constructed fairly obviously from the context, \eg by applying known transformations to existing ones, like converting singular to plural or incrementing a provided integer.
Whereas the former results provided a lower-bound on what is feasible, it is quite impossible to quantify precisely what the model ``could do", as the patterns learned by its millions of parameters can be highly complex. So we instead use the model's performance itself (studied in more depth in \cref{sec:model}) as an empirical datapoint: given that it is trained carefully and with ample capacity, we should expect that it provides at least evidence of this ability to produce correct new vocabulary. In contrast, we studied the trained model's accuracy on our 2,599 test samples; the patch introduced one or more tokens not present in the bug in over 75\% of the cases, yet the model predicts this new vocabulary only 5.6\% of the time. Worse, many of the ``new" tokens are not even entirely new; they may just constitute the addition of \verb+null+ check, which the model still does not anticipate. Even when (beam) searching across the top 25 most-probable sampled patches, the model only anticipates 14\% of the required new vocabulary. We stress that this is a well-trained model, which was able to achieve high accuracy on its training data and for which we used the most generalizable checkpoint after training for 100 epochs.
As such, machine translation models are already at a serious disadvantage here compared with NLP applications. This allows us to conclude our investigation of RQ1:

\begin{result}
The lack of information in the training data, vocabulary, and immediate context makes repairing as translation in its current form largely \textbf{infeasible}.
\end{result}

%Thus, we can answer our first research question:~\bray{??}

%\begin{result}
%Difference 1: The machine learner often has insufficient knowledge when generating program repairs as translation to reliably produce a bug-fixing patch.
%\end{result}

\vspace{-2mm}
\subsection{Architectural Design}
\label{subsec:architecture}

\begin{figure*}
  \begin{subfigure}{5.85cm}
        \includegraphics[width=5.75cm]{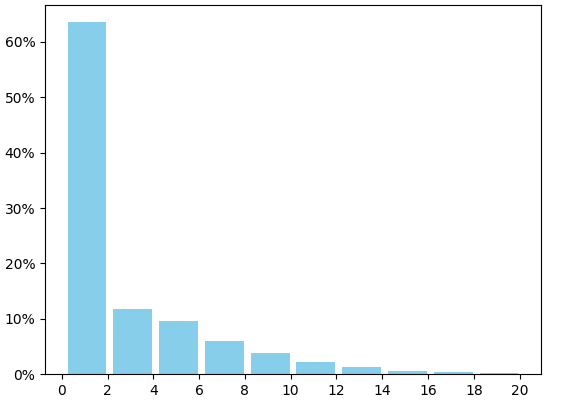}
  \caption{Edit Distance}
  \label{subfig:edit_distance}
  \vspace{-1mm}
  \end{subfigure}
  \begin{subfigure}{5.85cm}
        \includegraphics[width=5.85cm]{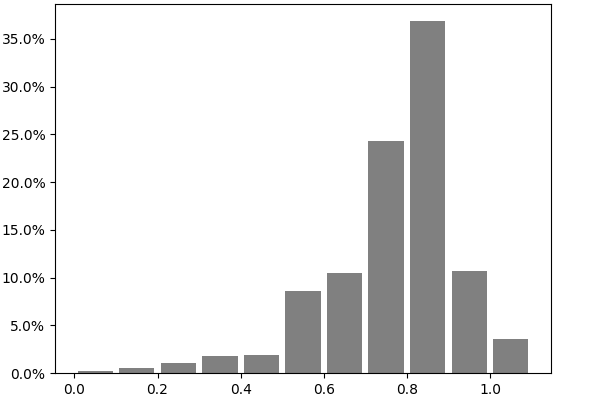}
  \caption{Intersection-over-union}
  \label{subfig:jaccard}
  \vspace{-1mm}
  \end{subfigure}
  \begin{subfigure}{5.85cm}
        \includegraphics[width=5.85cm]{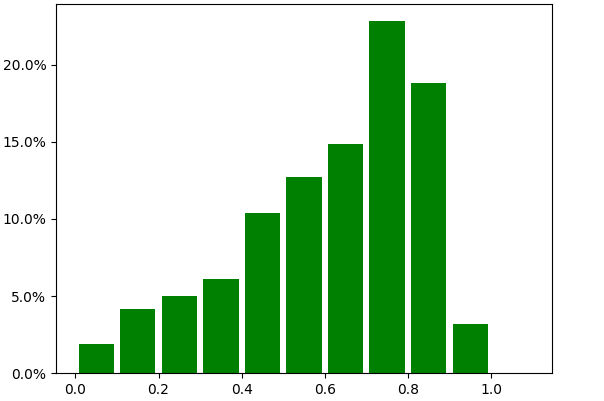}
  \caption{BLEU}
  \label{subfig:bleu}
  \vspace{-1mm}
  \end{subfigure}
\caption{\small Different Similarity Metrics regarding (bug, patch) pairs. X-axis of each histogram indicates the similarity score \wrt different metrics, and y-axis shows the ratio of samples within the corresponding range. The average edit distance between bugs and patches is 3.29. The edit distances of 51.1\% samples are 1 and 63.6\% samples have an edit distance $\leq$ 2. The average intersection-over-union similarity is 0.76, and 88.4\% samples have a similarity $\geq$ 0.6. The average BLEU score is 0.61, and 72.5\% samples have a BLEU score $\geq$ 0.5.}
\label{fig:bug_patch_similarity}
\vspace{-3mm}
\end{figure*}

Our second point of concern with translation models for program repair relates to the structural constraints assumed inherent in natural language generation: that text is auto-regressively produced left-to-right. This constraint is built into the translation model's (sequence-to-sequence) architecture and implies that a simple adoption for program repair requires the model to output the entire repair, producing the correct token at each point.

The flaw with this particular decision is different from the one in \Cref{subsec:semantics} in that it does not affect the \emph{feasibility} of the task (generating the entire repaired line is just as possible as \eg generating the change only). Instead, architectural mismatches between the model and the task impact the difficulty of training and the corresponding rate, and even the ultimate limit, of convergence on test data. This is because a) our models do not have infinite capacity and b) stochastic gradient descent is a local optimization; thus, these models tend to find a local minimum that matches the signal conveyed by the loss function. If this loss function prioritizes exact repetition of many tokens from the input, or a strong reliance on left-to-right production, this may negatively affect the actual quality (\eg overall accuracy) of the ultimate local minimum. In this section, we quantify this effect from the data statistics; in the next, we explore it further based on the model's convergent quality.

%\vspace{-2mm}
\subsubsection{The patch preserves most of the tokens in the bug}\footnote{This result applies to our study of small (one line) bug fixes; this may not hold for larger patches, which may be more likely to reconstruct the whole buggy module.}
\label{subsubsec:bug_patch_similar}
Bug-fixing modifications to committed code are often minor; the buggy line usually is already \emph{per se} a close approximation of the correct code,
with very subtle, minor flaws. To quantify this assertion, we first measure the similarity between real bugs and patches. We use three different metrics to evaluate the similarity of each (bug, patch) pair in our dataset, outlined in \cref{subsubsec: similarity_metrics}: token-level edit distance, (1-gram) intersection-over-union (which contributes a denominator to token-level overlap); and finally, mean BLEU-4 similarity to balance the overlap between tokens and sequences.

The results are shown in \Cref{fig:bug_patch_similarity}. The average edit distance for the samples in our dataset is $3.29$, but the distribution is long-tailed so this mean is somewhat inflated by the few large edits; the median distance is simply 1 -- $51.1\%$ of the samples only edit a single token to fix the bug, and $63.6\%$ of the samples have an edit distance up to 2. Thus, bug fixing modifications are often limited to just a select few tokens. \Cref{subfig:jaccard} further shows that bugs and patches share the majority of their vocabulary as well: the average Jaccard similarity is 0.76, and half the time the patch reserves more than $80\%$ of the bug's tokens.
This overlap extends to sequences of tokens as well: the mean BLEU score of a patch relative to its bug is 0.61. Two lines of code are considered very similar when their BLEU score is greater than 0.5, so bugs and their patches overlap strongly. For reference, the state-of-the-art results in NMT at this time are ca. 0.4, depending on the language pair. Program repair achieves far higher performance by simply copying the bug verbatim; yet, doing so would in no way approximate a \emph{good} repair.

This also confirms our intuition that bugs and patches are highly similar, and patches retain most tokens from the buggy version, rather than assembling code \emph{de novo}. This principally suggests that searching for the correct patch token by token, from left to right is a poor use of search space; a smart program repair tool should just predict which tokens are supposed to be preserved and focus on searching for the ones that require modifications. But is it really so bad to generate the entire patch; would copying the preserved tokens simply prove no concern for the models? We answer this question in the negative in \cref{subsec:objective}; first, we further analyze the \emph{types} of changes made in real repairs.

\vspace{-2mm}
\subsubsection{The patch tends to make minor changes to the bug's \underline{syntax}}
Grammars vary widely across languages. For example, subject-verb-object sequences (``I eat an apple") are abundant in English, but people seldom use them in verb-final languages like Tamil or Japanese. Because of this distinction, translating by merely substituting words in one language with another is often inappropriate. Instead, neural architectures capture the syntactic transformation between languages, as well as the translation of the underlying words. A recent study~\cite{ahmad2019difficulties} shows that the difference in word order among various languages is a significant feature that models learn, and \eg neural attention mechanisms are effective at this task. We are similarly curious whether this feature is prominent in the conversion of bugs to patches, as such information gives hints about how to adapt machine translation models appropriately. For example, given a buggy line: "\texttt{if (level >= damage - damage / 2)}" with patch: "\texttt{if (level <= damage - damage / 2)}" (a real sample from our dataset), we can see that the patch does not modify the syntax (in terms of the AST) of the bug, but only changes its semantics by changing the underlying tokens. We thus empirically study how often modifications that fix logical errors introduce changes to the syntactic structure of code.

\begin{table}[t]\centering
\vspace{-3mm}
\caption{\small Proportion of repairs in which the syntactic structures remains unchanged relative to bug, both for all samples in our training data and for those in which the patch both does and does not introduce novel tokens (relative to the bug).}
%\footnotesize
\small
\begin{tabular}{lr}\toprule
Setting & Proportion \\\midrule
All bugs &52.4\% \\
%Introduce unseen token but syntax unchanged & 56.2\% \\
Patches introducing new tokens & 56.2\% \\
Patches without new tokens & 20.6\% \\
\bottomrule
\end{tabular}
\label{tab: syntax_unchanged}
\vspace{-3.5mm}
\end{table}

Given a pair of bug and patch, we tokenize the code and use \textit{javalang} package \footnote{https://github.com/c2nes/javalang} to identify the syntactic type (\eg identifier, separator, integer) of each token. If a bug and patch have the exact same token type sequence, their syntactic structure is unchanged. \Cref{tab: syntax_unchanged} summarizes the resulting ratios. Slightly more than half of our patches preserve the exact syntactic structure of their bugs. Furthermore, the results are starkly different based on whether a patch introduces new tokens (relative to the bug; see \cref{subsubsec:new_vocab}); those that do even more rarely (56.2\%) change the syntax, while the other patches are often some kind of permutation of the bug's tokens that very rarely preserves syntactic ordering.

These observations have important implications. For one, they suggest that searching for unseen words across the entire vocabulary is rarely necessary; rather, the model could simply search for the tokens given a specific syntactic type~\cite{chakraborty2018codit}; \eg many patches replace just an operator to fix a bug. More generally, this suggests that the left-to-right generation process is thus not just inefficient, but all-but misdirected for such bugs: it requires the model to both copy a precise prefix, and then generate a single alternative from that context, where the original token was often already ``close" to being correct (\ie in the right syntactic ballpark). Patching such bugs, more than half of those in our dataset, in this way likely puts inordinate and unnecessary strain on the model, which we will quantify in the subsequent sections. First, we partially conclude our second research question:
\begin{result}
The machine translation architecture's generation process is a poor fit for program repair, which frequently retains most tokens from the bug while replacing just a few, and from a small candidate set.
\end{result}

\subsection{Program Repair via NMT (Objective)}
\label{subsec:objective}
Finally, when generating natural language translations, the goal is to correctly predict as many words of the target sentence as possible. The idea is that a translator that is likely to predict any one word given the input and previously predicted words (if any) is also likely to correctly generate the entire desired sentence by simply repeating this process until termination. Indeed, this tends to be quite accurate in general, in part because of the naturally auto-regressive, Markovian nature of text; a given prefix typically has only a small set of plausible continuations.

Given the observations in the preceding sections, this Markovian assumption seems precarious at best for program repair: the bug and repair often share a large, identical prefix (and suffix) that is then followed by incorrect tokens in the former and different, corrected ones in the latter. As such, we must question the validity of an objective function (both loss and metric) that values per-token prediction quality so strongly. Having said that, the aforementioned observations alone do not prove that there is a problem with this transplanted objective; the buggy token(s) may simply have been a particularly unnatural successor to its context \cite{Ray2016buggycode}, from which the corrected token(s) do, in fact, follow naturally. In this section, we empirically assess this concern.

Specifically, if the Markovian, auto-regressive objective used in natural language translation is a good fit for program repair as well, we would expect two things to be true:

\begin{enumerate}
    \item The per-token accuracy under auto-regressive teacher forcing correlates closely with the quality (\ie total accuracy) of the produced patch. That implies that the model correctly identifies the ``challenging" tokens, that need to be altered, as these dictate the overall correctness of the resulting patch.
    \item The model efficiently explores the repair space when sampling multiple patches (\eg using beam search). That implies that choosing the repair point by first copying tokens unaltered and then (auto-regressively) generating a different continuation is no distraction to the model.
\end{enumerate}

\begin{figure}[t]
    \centering
    \includegraphics[width=\linewidth]{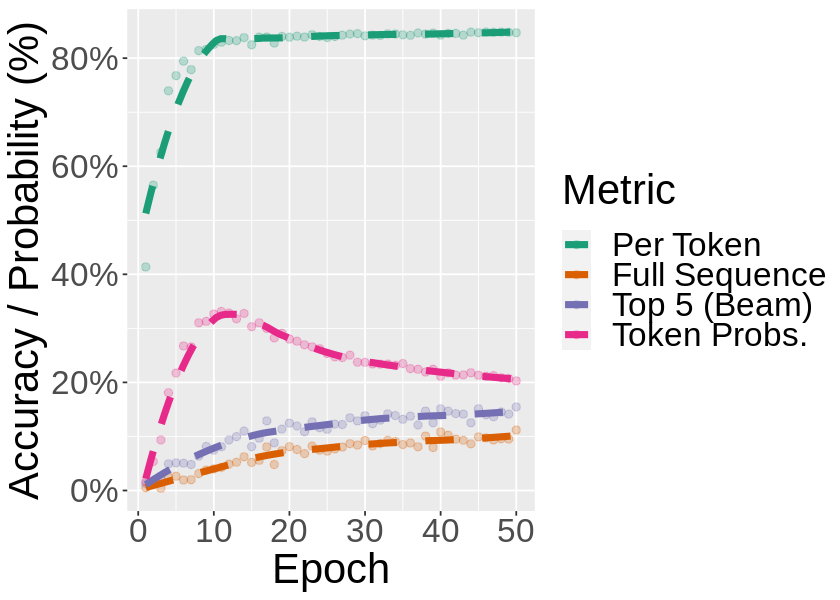}
    \caption{\small Performance trends (dashed line) and per-epoch results (points) on held-out data as training progresses, in terms of per-token accuracy subject to teacher forcing, accuracy at generating the complete sequence, and top-5 accuracy using beam search.}
    \label{fig:trainingcurve}
    \vspace{-4mm}
\end{figure}
We put both these expectations into the (empirical) test. \Cref{fig:trainingcurve} shows first the progression of various accuracies on our held-out data over the course of training. At the top, the teacher-forced token-level prediction accuracy increases steeply early on, throughout the first ca. 10 passes through the training data, but after that, it all-but plateaus. It does, in fact, still increase, but only very slightly after epoch 10 (from ca. 83\% to 85\%). This clearly shows two ``phases" (a bimodal pattern) in training this type of model: the model first trivially minimizes its loss (and thus achieves a high accuracy) through simple copying, but then struggles to match that strategy with predicting the correct change to achieve any more progress.

This initial copying translates into little real accuracy; the Full Sequence (\ie complete repair) prediction reaches just 4.5\% after 10 epochs, making nearly all its substantial progress afterwards. This has real training ramifications: we also visualize the progression of the per-token entropy (transformed to probabilities). In the first 10 epochs, the model quickly becomes very polarized, assigning high probability to the copied tokens; then it becomes clear that this yields very low probabilities for the few changed tokens, which entropy penalizes strongly. As a response, the model instead adopts a more balanced prediction to achieve higher overall repair quality.

To quantify the correlations in the face of this bimodality, a non-parametric (Spearman's rho) correlation test is in order. This does show that the two metrics (per-token accuracy and full sequence accuracy) are highly correlated ($\rho = 0.914$), even, though less so, after epoch 10 ($\rho = 0.863$). The latter result reflects that the remaining per-token accuracy increase translates into a disproportionally higher complete repair rate -- the missing 2\% token accuracy becomes ca. 7\% complete repair accuracy, nearly triple the levels at epoch 10. This implies a mixed answer to our first premise: the complete patch accuracy certainly follows the per-token accuracy, but the relation is far from direct and the latter is a highly misleading metric \emph{in ipso} due to its bimodal nature.

Finally, the figure shows that the odds of finding the correct patch in the top-5 generated samples is only a little higher than the top-1 prediction (ca. 5\% points at most); that gap actually shrinks as the top-1 prediction becomes more accurate, which suggests that the beam search finds few good novel/alternative patches. We would hope that, given the natural ambiguity in choosing the correct patch, the model learns to sample a diverse set of plausible corrections. Instead, from inspection of the generated samples, the model produces many very similar candidates, usually differing by just a few tokens. This too is likely an artifact of the training criteria, which prioritizes copying 80\% of the tokens over predicting the correct variation. Thus, we answer our final research question:
\begin{result}
The objective functions of NMT models are \textbf{inappropriate} for program repair, leading to reduced training efficacy on more appropriate metrics.
\end{result}
\section{Seq2seq model for program repair}
\label{sec:model}
As a fitting conclusion to our empirical and conceptual evaluation of the basic ``transplanted" approach to program repair as translation, it is appropriate to try and redesign the existing approach. This section demonstrates how observing and quantifying issues with an outside approach relates to principled and innovative modeling design: while observing concerns does not guarantee that improvements are straightforward (as we show in relation to context), it can improve performance by better relating the model to the task. We do this below, by eschewing past practice of trying to generate patch tokens directly, and instead generating \emph{edits}.

\subsection{Model Changes}
\begin{figure}[t]
    \centering
    \subcaptionbox{An edit-based repair model, which emits two pointers based on the encoder states that indicate the insertion start position and the removal end position. The decoder generates any missing tokens.}
    {\includegraphics[width=\linewidth, trim=0 370 100 0, clip]{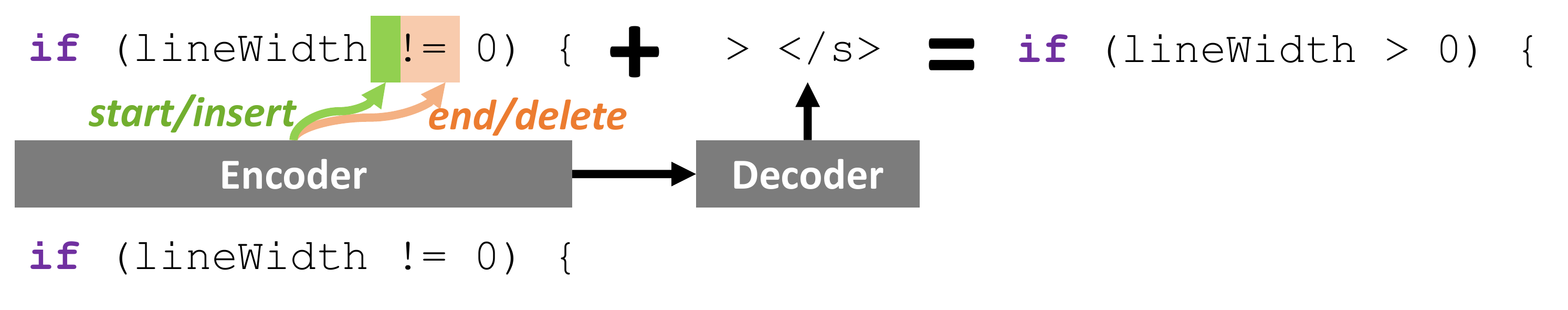}}\\ \vspace{.15cm}
    \subcaptionbox{Representation of a context-enriched repair model. The encoder functions as usual on a broader set of tokens; the decoder's attention is biased towards the highlighted (buggy) tokens.}
    {\includegraphics[width=\linewidth, trim=0 370 120 0, clip]{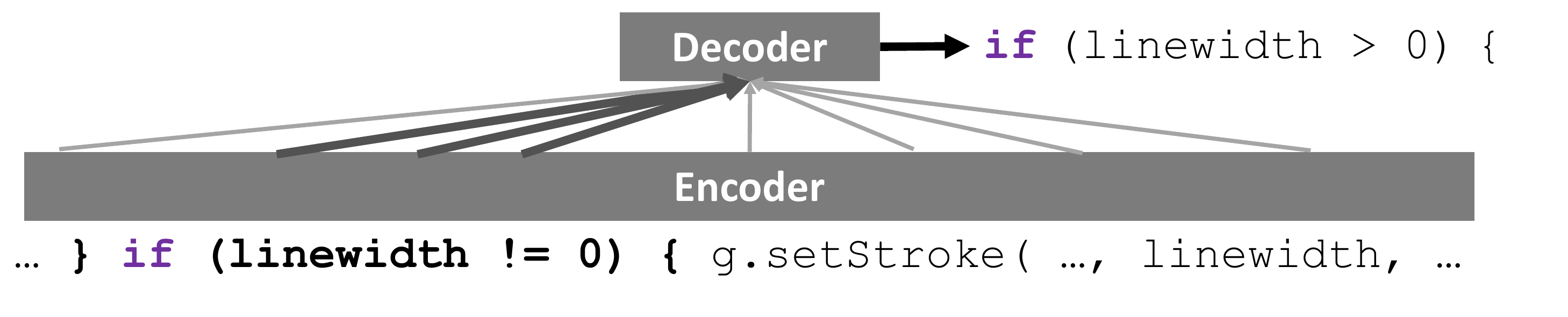}}
    \caption{\small Proposed architectural changes to the basic repair model on an example from our test data.}
    \label{fig:edit}
    \vspace{-4mm}
\end{figure}

We observed three main deficiencies with the existing translation approach: the inadequacy of relying on just the bug for enough information to produce a patch, the mismatch between typical repair actions and generating the entire corrected line, and the related divergence of training-objective, between per-token accuracy and whole-repair (both top-1 and beam-search) accuracy. Had we designed a machine learning approach for this problem from scratch, we would certainly attempt to incorporate both bug context, and a notion of repair \emph{edits} to reflect these aspects of program repair, as has also been proposed by some recent work \cite{tarlow2019graph2diff}. We propose to make both changes: \Cref{fig:edit} shows the two main architectural mechanisms we add to the base model to achieve this.

\vspace{-2mm}
\paragraph{Edits:} we model edits directly, as a  token-level ``diff" between the bug and patch. 
Our analysis of typical changes indicated that the bug and patch nearly always share a substantial prefix and/or suffix, with the repair occurring at some point in the middle of the line. We thus parse each bug/patch pair and find the longest overlapping prefix and suffix. Our model is augmented with two additional \emph{pointers} that correspond to insertion and deletion; the original decoder component (of the encoder-decoder architecture) 
is now pressed into service to output the diff (rather than directly generate the raw tokens in the fix).
There are three possible scenarios: \\
\textbf{No additions:} The prefix and suffix combined span the entire bug. This means that only tokens were added in the patch. In this case, the deletion pointer will just point to the start of the line, and the insertion pointer will indicate where the new tokens (which the decoder will emit) are to be added. \\
\textbf{No removals:} The prefix and suffix combined span the entire repair. In this case, only token deletions are needed, to go from the bug to the patch. So, the insertion and deletion pointer should correspond to the start and end of the segment to be deleted  within the buggy statement, and the decoder should just emit the ``</s>" termination symbol (an ``empty" patch). \\
\textbf{Additions \& Deletions:} A non-trivial change in both bug and patch. As a combination of the above, the two pointers should identify the segment to erase from the bug, while the decoder should generate all newly required tokens to insert instead.

\vspace{-1mm}
\paragraph{Context:} we also observed that the bug alone rarely provides enough (syntactic and semantic) information to reliably predict the necessary repair. The natural solution is to add a large amount of contextual tokens from the file containing the bug. Unfortunately, Transformers struggle to model very long sequences as their memory usage increases quadratically with sequence length. At the same time,  \cref{subsec:architecture} showed that even 20 lines of context is rarely enough to provide much missing vocabulary (which is itself only part of the information needed). We do not provide a new solution in this paper; rather, we empirically quantify the deficit from the model's perspective by adding up to 500 tokens of context and comparing the resulting performance. We ensure that the model is ``aware" of which tokens to repair by biasing the decoder's attention to the buggy tokens using the same biasing mechanism as in \cite{hellendoorn2020global}, in this case with a simple unary relation (\ie ``is part of the bug").

\vspace{-1mm}
\subsection{Results}
As \cref{tab:edit-results} shows, there are two main characteristics of the resulting models' performances. First, the edit-based enhancement clearly and substantially improves the accuracy over the baseline model, fixing an additional 22 bugs on our test set with its top prediction alone. Second, the contextual enhancement does not seem to help in its current form. We discuss both these results here.

\begin{table}[]
    \centering
    \caption{\small Repair accuracy on the (de-duplicated) test data of the various models that we propose in this paper.}
    \begin{tabular}{r | c | c | c}
        \textbf{Model} &  \textbf{Top-1} & \textbf{Top-5} & \textbf{Top-25} \\
        \toprule
        \emph{Baseline} & 3.30\% & 5.96\%  & \textbf{8.20\%} \\
        \emph{Edits} & \textbf{4.31\%}  & \textbf{6.14\%} & 7.83\%  \\
        \emph{Context} & 1.83\% & 3.57\% & 5.18\% \\
        \emph{Edits + Context} & 3.39\% & 4.08\% & 4.76\%\\
        \bottomrule
    \end{tabular}\\
    \label{tab:edit-results}
\end{table}

\begin{figure}[t]
    \centering
    \includegraphics[width=\linewidth]{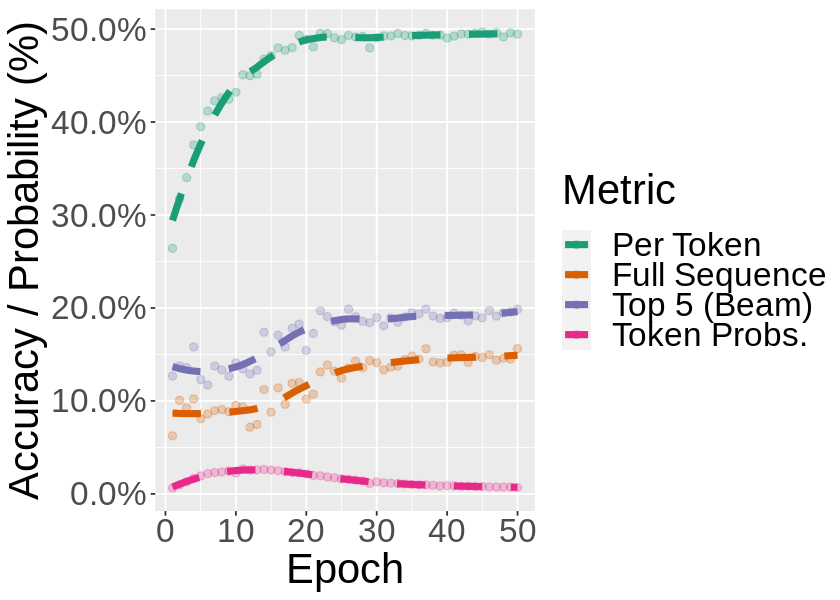}
    \caption{\small Performance of the edit-based model on held-out data as training progresses, in terms of per-token accuracy subject to teacher forcing, accuracy at generating the complete sequence, and top-5 accuracy using beam search.}
    \label{fig:trainingcurveedits}
    \vspace{-4mm}
\end{figure}

%\vspace{-2mm}
\paragraph{Edit model:} the edit-based model produces better-quality patches on our test data than the corresponding baseline. Its design is informed by our data analysis, and so it is arguably a better fit for this task. \Cref{fig:trainingcurveedits} shows its training behavior, to compare with that of the baseline model in \cref{fig:trainingcurve}; its ``per-token", teacher-forced accuracy increases much more smoothly\footnote{Their probability also displaying less of a ``spike" in early training.} and more in line with increases in the full repair prediction quality. It also displays a larger improvement in sampling accuracy between the top-1 and top-5 prediction, which remains consistently wide during training, suggesting that it better explores the search space with more diverse predictions.

Its design also allows the edit model to predict more newly introduced vocabulary in the patch relative to the bug; it does so 6.8\% and 15.7\% of the time (for the top-1 and top-25 samples respectively), compared to 5.6\% and 14\% of the base model. One notable difference is the gap between top-5 and top-25 sampling accuracy; the edit model is stronger in the former, but loses to the baseline in the latter. This appears to be due to the edit model having to commit to an insert \& delete pointer first, conditional upon which sampling is more bounded. To be clear, we did also sample these two pointers from their corresponding probability distributions and initialized the beam search with the 25 most probable different combinations of start and end pointers; but, in practice the model tended to choose a single pair with very high probability, so that it effectively only explored that set. This may be an interesting issue to pursue in future work.

\vspace{-1mm}
\paragraph{Context information:} the second missing element was the reliance on the bug alone as a source of patching information; in \cref{subsec:semantics}, we showed that the absence of context is an insurmountable obstacle that deprives the model of the necessary information to patch most bugs. However, identifying a problem and solving it are quite different things, as our results in \cref{tab:edit-results} show. Although we added a substantial amount of surrounding tokens (\ie 500) to the model's input, the resulting models' performance is quite poor, actually performing slightly worse than their context-free counterparts. This is likely due to the challenge of modeling large amounts of contextual information; although our models were trained to similar accuracies, they did so much slower and evidently with worse generalization.

This may point at several issues, but none seem quite responsible. For one, the attention mechanism we used may not adequately help the model locate the buggy bits; however, the model always emitted patches that were very similar to the bug. Similarly, the amount of context may simply be too little; \cref{tab:new_vocab} suggests that many useful tokens are only available far away from the bug. However, that table also implies that the immediate context \emph{should} help with ca 10-20\% of missing tokens, so this too does not explain the lack of performance. The model itself may simply have insufficient capacity to capture this much context, though we used a relatively large Transformer architecture and the model was trained to high accuracy. All this is to say that we do not know how to better integrate context in these models. This is not a bad thing; not all modeling improvements are obvious, but it is important that we understand the deficits first. Our empirical analysis helped us both identify it, and has laid a useful foundation for the kind of information to integrate in further improvements, even if it is not yet clear how.
\vspace{-2mm}
\section{Threats to Validity}
\label{sec:threats}
This paper presents a case-study of a specific type of program repair, which we explore in great empirical detail. As such, the main threats to the validity of our conclusion are external, relating to the generalization of our findings to both other types of defects and other model ``transplants" into SE research.

First, our data collection and analysis focused only on small, one-line fixes, since such bugs (and single-statement bugs) are both common and important, realistic target to current program repair models \cite{karampatsis2019singlestatement}. In addition, many existing NMT-based program-repair tools \cite{chen2018sequencer,lutellier2020coconut} are trained and tested on one-line or single-hunk bugs. As such, studying such bugs is both representative and impactful. Having said that, we do not claim, nor believe, that their empirical properties generalize to larger, more complex defects; these no doubt have their own non-trivial characteristics that deserve further investigation, especially if/when they become the subject of new models.

Secondly, we did not compare our model(s) \Cref{sec:model} with state-of-the-art, NMT-based, program repair tools. The goal of this work is not to present models with the best performance; rather, we are evaluating the feasibility of the general idea of ``patching as translation" using a general, representative modeling setup, especially in contrast to variations that depart from the translation metaphor. More broadly, there are many other cases of modeling transplants into our community, often with some alterations to fit the task; these may not all be harmful or mismatched, but they do all deserve careful empirical analysis to ensure that they achieve their potential efficacy in our community.
\section{Conclusion}
\label{conclusions}
In this work, we first present a comprehensive study to evaluate the conceit that "software patching is like the language translation" as a prototypical example of ``model transplant" from neighboring communities into SE. We empirically show that the translation paradigm does not capture bug-fixing very well for a range of reasons. We also use models themselves as empirical devices; we adapt the \emph{seq2seq} models used for translation to generate edits rather than raw tokens, which leads to promising improvements. 
We hope this work inspires more empirically-grounded research into  transplanting machine learning models to program repair, and other software engineering applications.
\section*{Acknowledgements}
This work is supported in part by NSF CCF-1414172, CCF-1845893, CNS-1842456, and CCF-1822965. Any opinions, findings, conclusions, or recommendations expressed herein are those of the authors, and do not necessarily reflect those of the US Government or NSF.

\balance
\bibliographystyle{ACM-Reference-Format}
\bibliography{reference}


\begin{thebibliography}{34}

%%% the Web address.

\ifx \showCODEN    \undefined \def \showCODEN     #1{\unskip}     \fi
\ifx \showDOI      \undefined \def \showDOI       #1{#1}\fi
\ifx \showISBNx    \undefined \def \showISBNx     #1{\unskip}     \fi
\ifx \showISBNxiii \undefined \def \showISBNxiii  #1{\unskip}     \fi
\ifx \showISSN     \undefined \def \showISSN      #1{\unskip}     \fi
\ifx \showLCCN     \undefined \def \showLCCN      #1{\unskip}     \fi
\ifx \shownote     \undefined \def \shownote      #1{#1}          \fi
\ifx \showarticletitle \undefined \def \showarticletitle #1{#1}   \fi
\ifx \showURL      \undefined \def \showURL       {\relax}        \fi
% invisible to TeX
\providecommand\bibfield[2]{#2}
\providecommand\bibinfo[2]{#2}
\providecommand\natexlab[1]{#1}
\providecommand\showeprint[2][]{arXiv:#2}

\bibitem[\protect\citeauthoryear{Ahmad, Zhang, Ma, Hovy, Chang, and Peng}{Ahmad
  et~al\mbox{.}}{2019}]%
        {ahmad2019difficulties}
\bibfield{author}{\bibinfo{person}{Wasi~Uddin Ahmad}, \bibinfo{person}{Zhisong
  Zhang}, \bibinfo{person}{Xuezhe Ma}, \bibinfo{person}{Eduard Hovy},
  \bibinfo{person}{Kai-Wei Chang}, {and} \bibinfo{person}{Nanyun Peng}.}
  \bibinfo{year}{2019}\natexlab{}.
\newblock \showarticletitle{On Difficulties of Cross-Lingual Transfer with
  Order Differences: A Case Study on Dependency Parsing}. In
  \bibinfo{booktitle}{\emph{NAACL}}.
\newblock


\bibitem[\protect\citeauthoryear{Allamanis, Brockschmidt, and
  Khademi}{Allamanis et~al\mbox{.}}{2018}]%
        {allamanis18learning}
\bibfield{author}{\bibinfo{person}{Miltiadis Allamanis}, \bibinfo{person}{Marc
  Brockschmidt}, {and} \bibinfo{person}{Mahmoud Khademi}.}
  \bibinfo{year}{2018}\natexlab{}.
\newblock \showarticletitle{Learning to Represent Programs with Graphs}. In
  \bibinfo{booktitle}{\emph{International Conference on Learning
  Representations (ICLR)}}.
\newblock


\bibitem[\protect\citeauthoryear{Alon, Zilberstein, Levy, and Yahav}{Alon
  et~al\mbox{.}}{2019}]%
        {Alon2019code2vec}
\bibfield{author}{\bibinfo{person}{Uri Alon}, \bibinfo{person}{Meital
  Zilberstein}, \bibinfo{person}{Omer Levy}, {and} \bibinfo{person}{Eran
  Yahav}.} \bibinfo{year}{2019}\natexlab{}.
\newblock \showarticletitle{Code2vec: Learning Distributed Representations of
  Code}.
\newblock \bibinfo{journal}{\emph{Proc. ACM Program. Lang.}}
  \bibinfo{volume}{3}, \bibinfo{number}{POPL}, Article \bibinfo{articleno}{40}
  (\bibinfo{date}{Jan.} \bibinfo{year}{2019}), \bibinfo{numpages}{29}~pages.
\newblock
\urldef\tempurl%
\url{https://doi.org/10.1145/3290353}
\showDOI{\tempurl}


\bibitem[\protect\citeauthoryear{Bahdanau, Cho, and Bengio}{Bahdanau
  et~al\mbox{.}}{2015}]%
        {Bahdanau2015NeuralMT}
\bibfield{author}{\bibinfo{person}{Dzmitry Bahdanau},
  \bibinfo{person}{Kyunghyun Cho}, {and} \bibinfo{person}{Yoshua Bengio}.}
  \bibinfo{year}{2015}\natexlab{}.
\newblock \showarticletitle{Neural Machine Translation by Jointly Learning to
  Align and Translate}.
\newblock \bibinfo{journal}{\emph{CoRR}}  \bibinfo{volume}{abs/1409.0473}
  (\bibinfo{year}{2015}).
\newblock


\bibitem[\protect\citeauthoryear{Chakraborty, Ding, Allamanis, and
  Ray}{Chakraborty et~al\mbox{.}}{2018}]%
        {chakraborty2018codit}
\bibfield{author}{\bibinfo{person}{Saikat Chakraborty},
  \bibinfo{person}{Yangruibo Ding}, \bibinfo{person}{Miltiadis Allamanis},
  {and} \bibinfo{person}{Baishakhi Ray}.} \bibinfo{year}{2018}\natexlab{}.
\newblock \bibinfo{title}{CODIT: Code Editing with Tree-Based Neural Models}.
\newblock
\newblock
\showeprint[arxiv]{cs.SE/1810.00314}


\bibitem[\protect\citeauthoryear{Chen, Kommrusch, Tufano, Pouchet, Poshyvanyk,
  and Monperrus}{Chen et~al\mbox{.}}{2019}]%
        {chen2018sequencer}
\bibfield{author}{\bibinfo{person}{Zimin Chen}, \bibinfo{person}{Steve
  Kommrusch}, \bibinfo{person}{Michele Tufano}, \bibinfo{person}{Louis-No{\"e}l
  Pouchet}, \bibinfo{person}{Denys Poshyvanyk}, {and} \bibinfo{person}{Martin
  Monperrus}.} \bibinfo{year}{2019}\natexlab{}.
\newblock \showarticletitle{SequenceR: Sequence-to-Sequence Learning for
  End-to-End Program Repair}.
\newblock \bibinfo{journal}{\emph{IEEE Transaction on Software Engineering}}
  (\bibinfo{year}{2019}).
\newblock


\bibitem[\protect\citeauthoryear{Cho, van Merrienboer, Bahdanau, and
  Bengio}{Cho et~al\mbox{.}}{2014}]%
        {Cho2014OnTP}
\bibfield{author}{\bibinfo{person}{Kyunghyun Cho}, \bibinfo{person}{Bart van
  Merrienboer}, \bibinfo{person}{Dzmitry Bahdanau}, {and}
  \bibinfo{person}{Yoshua Bengio}.} \bibinfo{year}{2014}\natexlab{}.
\newblock \showarticletitle{On the Properties of Neural Machine Translation:
  Encoder-Decoder Approaches}.
\newblock \bibinfo{journal}{\emph{ArXiv}}  \bibinfo{volume}{abs/1409.1259}
  (\bibinfo{year}{2014}).
\newblock


\bibitem[\protect\citeauthoryear{Dinella, Dai, Li, Naik, Song, and
  Wang}{Dinella et~al\mbox{.}}{2020}]%
        {Dinella2020HoppityLG}
\bibfield{author}{\bibinfo{person}{Elizabeth Dinella}, \bibinfo{person}{Hanjun
  Dai}, \bibinfo{person}{Ziyang Li}, \bibinfo{person}{M. Naik},
  \bibinfo{person}{L. Song}, {and} \bibinfo{person}{K. Wang}.}
  \bibinfo{year}{2020}\natexlab{}.
\newblock \showarticletitle{Hoppity: Learning Graph Transformations to Detect
  and Fix Bugs in Programs}. In \bibinfo{booktitle}{\emph{ICLR}}.
\newblock


\bibitem[\protect\citeauthoryear{Goues, Nguyen, Forrest, and Weimer}{Goues
  et~al\mbox{.}}{2012}]%
        {Goues2012GenProgAG}
\bibfield{author}{\bibinfo{person}{Claire~Le Goues}, \bibinfo{person}{ThanhVu
  Nguyen}, \bibinfo{person}{Stephanie Forrest}, {and} \bibinfo{person}{Westley
  Weimer}.} \bibinfo{year}{2012}\natexlab{}.
\newblock \showarticletitle{GenProg: A Generic Method for Automatic Software
  Repair}.
\newblock \bibinfo{journal}{\emph{IEEE Transactions on Software Engineering}}
  \bibinfo{volume}{38} (\bibinfo{year}{2012}), \bibinfo{pages}{54--72}.
\newblock


\bibitem[\protect\citeauthoryear{Gu, Zhang, Zhang, and Kim}{Gu
  et~al\mbox{.}}{2017}]%
        {gu2017deepam}
\bibfield{author}{\bibinfo{person}{Xiaodong Gu}, \bibinfo{person}{Hongyu
  Zhang}, \bibinfo{person}{Dongmei Zhang}, {and} \bibinfo{person}{Sunghun
  Kim}.} \bibinfo{year}{2017}\natexlab{}.
\newblock \showarticletitle{DeepAM: Migrate APIs with Multi-Modal Sequence to
  Sequence Learning}. In \bibinfo{booktitle}{\emph{Proceedings of the 26th
  International Joint Conference on Artificial Intelligence}} (Melbourne,
  Australia) \emph{(\bibinfo{series}{IJCAI '17})}. \bibinfo{publisher}{AAAI
  Press}, \bibinfo{pages}{3675--3681}.
\newblock
\showISBNx{9780999241103}


\bibitem[\protect\citeauthoryear{Gupta, Pal, Kanade, and Shevade}{Gupta
  et~al\mbox{.}}{2017}]%
        {gupta2017deepfix}
\bibfield{author}{\bibinfo{person}{Rahul Gupta}, \bibinfo{person}{Soham Pal},
  \bibinfo{person}{Aditya Kanade}, {and} \bibinfo{person}{Shirish Shevade}.}
  \bibinfo{year}{2017}\natexlab{}.
\newblock \showarticletitle{{DeepFix}: Fixing Common {C} Language Errors by
  Deep Learning.}. In \bibinfo{booktitle}{\emph{AAAI}}.
  \bibinfo{pages}{1345--1351}.
\newblock


\bibitem[\protect\citeauthoryear{Hellendoorn and Devanbu}{Hellendoorn and
  Devanbu}{2017}]%
        {hellendoorn2017deep}
\bibfield{author}{\bibinfo{person}{Vincent~J. Hellendoorn} {and}
  \bibinfo{person}{Premkumar Devanbu}.} \bibinfo{year}{2017}\natexlab{}.
\newblock \showarticletitle{Are deep neural networks the best choice for
  modeling source code?}. In \bibinfo{booktitle}{\emph{Proceedings of the 2017
  11th Joint Meeting on Foundations of Software Engineering}}. ACM,
  \bibinfo{pages}{763--773}.
\newblock


\bibitem[\protect\citeauthoryear{Hellendoorn, Maniatis, Singh, Sutton, and
  Bieber}{Hellendoorn et~al\mbox{.}}{2020}]%
        {hellendoorn2020global}
\bibfield{author}{\bibinfo{person}{Vincent~J Hellendoorn},
  \bibinfo{person}{Petros Maniatis}, \bibinfo{person}{Rishabh Singh},
  \bibinfo{person}{Charles Sutton}, {and} \bibinfo{person}{David Bieber}.}
  \bibinfo{year}{2020}\natexlab{}.
\newblock \showarticletitle{Global relational models of source code}. In
  \bibinfo{booktitle}{\emph{2020 8th International Conference on Learning
  Representations (ICLR)}}.
\newblock


\bibitem[\protect\citeauthoryear{Hindle, Barr, Su, Gabel, and Devanbu}{Hindle
  et~al\mbox{.}}{2012}]%
        {Hindle2012naturalness}
\bibfield{author}{\bibinfo{person}{Abram Hindle}, \bibinfo{person}{Earl~T.
  Barr}, \bibinfo{person}{Zhendong Su}, \bibinfo{person}{Mark Gabel}, {and}
  \bibinfo{person}{Premkumar Devanbu}.} \bibinfo{year}{2012}\natexlab{}.
\newblock \showarticletitle{On the Naturalness of Software}. In
  \bibinfo{booktitle}{\emph{Proceedings of the 34th International Conference on
  Software Engineering}} (Zurich, Switzerland) \emph{(\bibinfo{series}{ICSE
  '12})}. \bibinfo{publisher}{IEEE Press}, \bibinfo{pages}{837--847}.
\newblock
\showISBNx{9781467310673}


\bibitem[\protect\citeauthoryear{Hochreiter and Schmidhuber}{Hochreiter and
  Schmidhuber}{1997}]%
        {hochreiter_1997_lstm}
\bibfield{author}{\bibinfo{person}{Sepp Hochreiter} {and}
  \bibinfo{person}{J\"{u}rgen Schmidhuber}.} \bibinfo{year}{1997}\natexlab{}.
\newblock \showarticletitle{Long Short-Term Memory}.
\newblock \bibinfo{journal}{\emph{Neural Comput.}} \bibinfo{volume}{9},
  \bibinfo{number}{8} (\bibinfo{date}{Nov.} \bibinfo{year}{1997}),
  \bibinfo{pages}{1735--1780}.
\newblock
\showISSN{0899-7667}
\urldef\tempurl%
\url{https://doi.org/10.1162/neco.1997.9.8.1735}
\showDOI{\tempurl}


\bibitem[\protect\citeauthoryear{Iyer, Konstas, Cheung, and Zettlemoyer}{Iyer
  et~al\mbox{.}}{2016}]%
        {iyer2016summarizing}
\bibfield{author}{\bibinfo{person}{Srinivasan Iyer}, \bibinfo{person}{Ioannis
  Konstas}, \bibinfo{person}{Alvin Cheung}, {and} \bibinfo{person}{Luke
  Zettlemoyer}.} \bibinfo{year}{2016}\natexlab{}.
\newblock \showarticletitle{Summarizing Source Code using a Neural Attention
  Model}. In \bibinfo{booktitle}{\emph{Proceedings of the 54th Annual Meeting
  of the Association for Computational Linguistics (Volume 1: Long Papers)}}.
  \bibinfo{publisher}{Association for Computational Linguistics},
  \bibinfo{address}{Berlin, Germany}, \bibinfo{pages}{2073--2083}.
\newblock
\urldef\tempurl%
\url{https://doi.org/10.18653/v1/P16-1195}
\showDOI{\tempurl}


\bibitem[\protect\citeauthoryear{Jiang, Xiong, Zhang, Gao, and Chen}{Jiang
  et~al\mbox{.}}{2018}]%
        {Simfix:2018}
\bibfield{author}{\bibinfo{person}{Jiajun Jiang}, \bibinfo{person}{Yingfei
  Xiong}, \bibinfo{person}{Hongyu Zhang}, \bibinfo{person}{Qing Gao}, {and}
  \bibinfo{person}{Xiangqun Chen}.} \bibinfo{year}{2018}\natexlab{}.
\newblock \showarticletitle{Shaping Program Repair Space with Existing Patches
  and Similar Code} \emph{(\bibinfo{series}{ISSTA})}.
\newblock
\urldef\tempurl%
\url{https://doi.org/10.1145/3213846.3213871}
\showDOI{\tempurl}


\bibitem[\protect\citeauthoryear{Karampatsis, Babii, Robbes, Sutton, and
  Janes}{Karampatsis et~al\mbox{.}}{2020}]%
        {karampatsis2020big}
\bibfield{author}{\bibinfo{person}{Rafael-Michael Karampatsis},
  \bibinfo{person}{Hlib Babii}, \bibinfo{person}{Romain Robbes},
  \bibinfo{person}{Charles Sutton}, {and} \bibinfo{person}{Andrea Janes}.}
  \bibinfo{year}{2020}\natexlab{}.
\newblock \bibinfo{title}{Big Code != Big Vocabulary: Open-Vocabulary Models
  for Source Code}.
\newblock
\newblock
\showeprint[arxiv]{cs.SE/2003.07914}


\bibitem[\protect\citeauthoryear{Karampatsis and Sutton}{Karampatsis and
  Sutton}{2019}]%
        {karampatsis2019singlestatement}
\bibfield{author}{\bibinfo{person}{Rafael-Michael Karampatsis} {and}
  \bibinfo{person}{Charles Sutton}.} \bibinfo{year}{2019}\natexlab{}.
\newblock \bibinfo{title}{How Often Do Single-Statement Bugs Occur? The
  ManySStuBs4J Dataset}.
\newblock
\newblock
\showeprint[arxiv]{cs.SE/1905.13334}


\bibitem[\protect\citeauthoryear{Le, Chu, Lo, Le~Goues, and Visser}{Le
  et~al\mbox{.}}{2017}]%
        {le_2017_s3}
\bibfield{author}{\bibinfo{person}{Xuan-Bach~D. Le}, \bibinfo{person}{Duc-Hiep
  Chu}, \bibinfo{person}{David Lo}, \bibinfo{person}{Claire Le~Goues}, {and}
  \bibinfo{person}{Willem Visser}.} \bibinfo{year}{2017}\natexlab{}.
\newblock \showarticletitle{S3: Syntax- and Semantic-Guided Repair Synthesis
  via Programming by Examples}. In \bibinfo{booktitle}{\emph{Proceedings of the
  2017 11th Joint Meeting on Foundations of Software Engineering}} (Paderborn,
  Germany) \emph{(\bibinfo{series}{ESEC/FSE 2017})}.
  \bibinfo{publisher}{Association for Computing Machinery},
  \bibinfo{address}{New York, NY, USA}, \bibinfo{pages}{593--604}.
\newblock
\showISBNx{9781450351058}
\urldef\tempurl%
\url{https://doi.org/10.1145/3106237.3106309}
\showDOI{\tempurl}


\bibitem[\protect\citeauthoryear{Long and Rinard}{Long and Rinard}{2016}]%
        {fan2016analysis}
\bibfield{author}{\bibinfo{person}{Fan Long} {and} \bibinfo{person}{Martin
  Rinard}.} \bibinfo{year}{2016}\natexlab{}.
\newblock \showarticletitle{An Analysis of the Search Spaces for Generate and
  Validate Patch Generation Systems}. In \bibinfo{booktitle}{\emph{2016
  IEEE/ACM 38th International Conference on Software Engineering (ICSE)}}.
\newblock


\bibitem[\protect\citeauthoryear{Lutellier, Pham, Pang, Li, Wei, and
  Tan}{Lutellier et~al\mbox{.}}{2020}]%
        {lutellier2020coconut}
\bibfield{author}{\bibinfo{person}{Thibaud Lutellier},
  \bibinfo{person}{Hung~Viet Pham}, \bibinfo{person}{Lawrence Pang},
  \bibinfo{person}{Yitong Li}, \bibinfo{person}{Moshi Wei}, {and}
  \bibinfo{person}{Lin Tan}.} \bibinfo{year}{2020}\natexlab{}.
\newblock \showarticletitle{CoCoNuT: Combining Context-Aware Neural Translation
  Models Using Ensemble for Program Repair} \emph{(\bibinfo{series}{ISSTA
  2020})}. \bibinfo{publisher}{Association for Computing Machinery},
  \bibinfo{address}{New York, NY, USA}, \bibinfo{pages}{101--114}.
\newblock
\showISBNx{9781450380089}
\urldef\tempurl%
\url{https://doi.org/10.1145/3395363.3397369}
\showDOI{\tempurl}


\bibitem[\protect\citeauthoryear{Papineni, Roukos, Ward, and Zhu}{Papineni
  et~al\mbox{.}}{2002}]%
        {papineni_02_bleu}
\bibfield{author}{\bibinfo{person}{Kishore Papineni}, \bibinfo{person}{Salim
  Roukos}, \bibinfo{person}{Todd Ward}, {and} \bibinfo{person}{Wei-Jing Zhu}.}
  \bibinfo{year}{2002}\natexlab{}.
\newblock \showarticletitle{BLEU: A Method for Automatic Evaluation of Machine
  Translation}. In \bibinfo{booktitle}{\emph{Proceedings of the 40th Annual
  Meeting on Association for Computational Linguistics}} (Philadelphia,
  Pennsylvania) \emph{(\bibinfo{series}{ACL '02})}.
  \bibinfo{publisher}{Association for Computational Linguistics},
  \bibinfo{address}{USA}, \bibinfo{pages}{311--318}.
\newblock
\urldef\tempurl%
\url{https://doi.org/10.3115/1073083.1073135}
\showDOI{\tempurl}


\bibitem[\protect\citeauthoryear{Qi, Mao, Lei, Dai, and Wang}{Qi
  et~al\mbox{.}}{2014}]%
        {qi_2014_rsrepair}
\bibfield{author}{\bibinfo{person}{Yuhua Qi}, \bibinfo{person}{Xiaoguang Mao},
  \bibinfo{person}{Yan Lei}, \bibinfo{person}{Ziying Dai}, {and}
  \bibinfo{person}{Chengsong Wang}.} \bibinfo{year}{2014}\natexlab{}.
\newblock \showarticletitle{The Strength of Random Search on Automated Program
  Repair}. In \bibinfo{booktitle}{\emph{Proceedings of the 36th International
  Conference on Software Engineering}} (Hyderabad, India)
  \emph{(\bibinfo{series}{ICSE '14})}. \bibinfo{publisher}{Association for
  Computing Machinery}, \bibinfo{address}{New York, NY, USA},
  \bibinfo{pages}{254--265}.
\newblock
\showISBNx{9781450327565}
\urldef\tempurl%
\url{https://doi.org/10.1145/2568225.2568254}
\showDOI{\tempurl}


\bibitem[\protect\citeauthoryear{Qi, Long, Achour, and Rinard}{Qi
  et~al\mbox{.}}{2015}]%
        {qi_2015_kali}
\bibfield{author}{\bibinfo{person}{Zichao Qi}, \bibinfo{person}{Fan Long},
  \bibinfo{person}{Sara Achour}, {and} \bibinfo{person}{Martin Rinard}.}
  \bibinfo{year}{2015}\natexlab{}.
\newblock \showarticletitle{An Analysis of Patch Plausibility and Correctness
  for Generate-and-Validate Patch Generation Systems}. In
  \bibinfo{booktitle}{\emph{Proceedings of the 2015 International Symposium on
  Software Testing and Analysis}} (Baltimore, MD, USA)
  \emph{(\bibinfo{series}{ISSTA '15})}. \bibinfo{publisher}{Association for
  Computing Machinery}, \bibinfo{address}{New York, NY, USA},
  \bibinfo{pages}{24--36}.
\newblock
\showISBNx{9781450336208}
\urldef\tempurl%
\url{https://doi.org/10.1145/2771783.2771791}
\showDOI{\tempurl}


\bibitem[\protect\citeauthoryear{Ray, Hellendoorn, Godhane, Tu, Bacchelli, and
  Devanbu}{Ray et~al\mbox{.}}{2016}]%
        {Ray2016buggycode}
\bibfield{author}{\bibinfo{person}{Baishakhi Ray}, \bibinfo{person}{Vincent
  Hellendoorn}, \bibinfo{person}{Saheel Godhane}, \bibinfo{person}{Zhaopeng
  Tu}, \bibinfo{person}{Alberto Bacchelli}, {and} \bibinfo{person}{Premkumar
  Devanbu}.} \bibinfo{year}{2016}\natexlab{}.
\newblock \showarticletitle{On the "Naturalness" of Buggy Code}. In
  \bibinfo{booktitle}{\emph{Proceedings of the 38th International Conference on
  Software Engineering}} (Austin, Texas) \emph{(\bibinfo{series}{ICSE '16})}.
  \bibinfo{publisher}{Association for Computing Machinery},
  \bibinfo{address}{New York, NY, USA}, \bibinfo{pages}{428--439}.
\newblock
\showISBNx{9781450339001}
\urldef\tempurl%
\url{https://doi.org/10.1145/2884781.2884848}
\showDOI{\tempurl}


\bibitem[\protect\citeauthoryear{Sennrich, Haddow, and Birch}{Sennrich
  et~al\mbox{.}}{2016}]%
        {sennrich-etal-2016-neural}
\bibfield{author}{\bibinfo{person}{Rico Sennrich}, \bibinfo{person}{Barry
  Haddow}, {and} \bibinfo{person}{Alexandra Birch}.}
  \bibinfo{year}{2016}\natexlab{}.
\newblock \showarticletitle{Neural Machine Translation of Rare Words with
  Subword Units}. In \bibinfo{booktitle}{\emph{Proceedings of the 54th Annual
  Meeting of the Association for Computational Linguistics (Volume 1: Long
  Papers)}}. \bibinfo{publisher}{Association for Computational Linguistics},
  \bibinfo{address}{Berlin, Germany}, \bibinfo{pages}{1715--1725}.
\newblock
\urldef\tempurl%
\url{https://doi.org/10.18653/v1/P16-1162}
\showDOI{\tempurl}


\bibitem[\protect\citeauthoryear{Sutskever, Vinyals, and Le}{Sutskever
  et~al\mbox{.}}{2014}]%
        {sutskever2014seq2seq}
\bibfield{author}{\bibinfo{person}{Ilya Sutskever}, \bibinfo{person}{Oriol
  Vinyals}, {and} \bibinfo{person}{Quoc~V. Le}.}
  \bibinfo{year}{2014}\natexlab{}.
\newblock \showarticletitle{Sequence to Sequence Learning with Neural
  Networks}. In \bibinfo{booktitle}{\emph{Proceedings of the 27th International
  Conference on Neural Information Processing Systems - Volume 2}} (Montreal,
  Canada) \emph{(\bibinfo{series}{NIPS '14})}. \bibinfo{publisher}{MIT Press},
  \bibinfo{address}{Cambridge, MA, USA}, \bibinfo{pages}{3104--3112}.
\newblock


\bibitem[\protect\citeauthoryear{Tarlow, Moitra, Rice, Chen, Manzagol, Sutton,
  and Aftandilian}{Tarlow et~al\mbox{.}}{2019}]%
        {tarlow2019graph2diff}
\bibfield{author}{\bibinfo{person}{Daniel Tarlow}, \bibinfo{person}{Subhodeep
  Moitra}, \bibinfo{person}{Andrew Rice}, \bibinfo{person}{Zimin Chen},
  \bibinfo{person}{Pierre-Antoine Manzagol}, \bibinfo{person}{Charles Sutton},
  {and} \bibinfo{person}{Edward Aftandilian}.} \bibinfo{year}{2019}\natexlab{}.
\newblock \showarticletitle{Learning to Fix Build Errors with Graph2Diff Neural
  Networks}.
\newblock \bibinfo{journal}{\emph{arXiv preprint arXiv:1911.01205}}
  (\bibinfo{year}{2019}).
\newblock


\bibitem[\protect\citeauthoryear{Tufano, Watson, Bavota, Di~Penta, White, and
  Poshyvanyk}{Tufano et~al\mbox{.}}{2018}]%
        {tufano2018nmt_bug_fix}
\bibfield{author}{\bibinfo{person}{Michele Tufano}, \bibinfo{person}{Cody
  Watson}, \bibinfo{person}{Gabriele Bavota}, \bibinfo{person}{Massimiliano
  Di~Penta}, \bibinfo{person}{Martin White}, {and} \bibinfo{person}{Denys
  Poshyvanyk}.} \bibinfo{year}{2018}\natexlab{}.
\newblock \showarticletitle{An Empirical Investigation into Learning Bug-Fixing
  Patches in the Wild via Neural Machine Translation}. In
  \bibinfo{booktitle}{\emph{Proceedings of the 2018 33rd ACM/IEEE International
  Conference on Automated Software Engineering}}.
\newblock


\bibitem[\protect\citeauthoryear{Vasic, Kanade, Maniatis, Bieber, and
  Rishabh}{Vasic et~al\mbox{.}}{2019}]%
        {vasic2019pointernet}
\bibfield{author}{\bibinfo{person}{Marko Vasic}, \bibinfo{person}{Aditya
  Kanade}, \bibinfo{person}{Petros Maniatis}, \bibinfo{person}{David Bieber},
  {and} \bibinfo{person}{Singh Rishabh}.} \bibinfo{year}{2019}\natexlab{}.
\newblock \showarticletitle{Neural Program Repair by Jointly Learning to
  Localize and Repair.}. In \bibinfo{booktitle}{\emph{ICLR}}.
\newblock


\bibitem[\protect\citeauthoryear{Vaswani, Shazeer, Parmar, Uszkoreit, Jones,
  Gomez, Kaiser, and Polosukhin}{Vaswani et~al\mbox{.}}{2017}]%
        {vaswani2017transformer}
\bibfield{author}{\bibinfo{person}{Ashish Vaswani}, \bibinfo{person}{Noam
  Shazeer}, \bibinfo{person}{Niki Parmar}, \bibinfo{person}{Jakob Uszkoreit},
  \bibinfo{person}{Llion Jones}, \bibinfo{person}{Aidan~N. Gomez},
  \bibinfo{person}{Lukasz Kaiser}, {and} \bibinfo{person}{Illia Polosukhin}.}
  \bibinfo{year}{2017}\natexlab{}.
\newblock \showarticletitle{Attention is All You Need}. In
  \bibinfo{booktitle}{\emph{Proceedings of the 31st International Conference on
  Neural Information Processing Systems}} (Long Beach, California, USA)
  \emph{(\bibinfo{series}{NIPS '17})}. \bibinfo{publisher}{Curran Associates
  Inc.}, \bibinfo{address}{Red Hook, NY, USA}, \bibinfo{pages}{6000--6010}.
\newblock
\showISBNx{9781510860964}


\bibitem[\protect\citeauthoryear{Vinyals, Fortunato, and Jaitly}{Vinyals
  et~al\mbox{.}}{2015}]%
        {vinyals2015pointernet}
\bibfield{author}{\bibinfo{person}{Oriol Vinyals}, \bibinfo{person}{Meire
  Fortunato}, {and} \bibinfo{person}{Navdeep Jaitly}.}
  \bibinfo{year}{2015}\natexlab{}.
\newblock \showarticletitle{Pointer network.}. In
  \bibinfo{booktitle}{\emph{Proceedings of the 28th International Conference on
  Neural Information Processing Systems (NIPS '15)}}.
  \bibinfo{pages}{2692--2700}.
\newblock


\bibitem[\protect\citeauthoryear{Zhang, Wang, Zhang, Sun, Wang, and Liu}{Zhang
  et~al\mbox{.}}{2019}]%
        {zhang2019astnn}
\bibfield{author}{\bibinfo{person}{Jian Zhang}, \bibinfo{person}{Xu Wang},
  \bibinfo{person}{Hongyu Zhang}, \bibinfo{person}{Hailong Sun},
  \bibinfo{person}{Kaixuan Wang}, {and} \bibinfo{person}{Xudong Liu}.}
  \bibinfo{year}{2019}\natexlab{}.
\newblock \showarticletitle{A Novel Neural Source Code Representation Based on
  Abstract Syntax Tree}. In \bibinfo{booktitle}{\emph{Proceedings of the 41st
  International Conference on Software Engineering}} (Montreal, Quebec, Canada)
  \emph{(\bibinfo{series}{ICSE '19})}. \bibinfo{publisher}{IEEE Press},
  \bibinfo{pages}{783--794}.
\newblock
\urldef\tempurl%
\url{https://doi.org/10.1109/ICSE.2019.00086}
\showDOI{\tempurl}


\end{thebibliography}

% \appendix

\end{document}